\documentclass{article}
\usepackage{graphicx} 
\usepackage{graphicx,amsmath,amssymb,tikz,amsthm} 
\usepackage{arydshln}
\usepackage{mathtools}
\usepackage{bm}
\usepackage[linktocpage=true,
  colorlinks=true, 
  pdfborder={0 0 0},
  linkcolor=blue,
  citecolor=red,
  filecolor=yellow,
  urlcolor=blue,
  bookmarks,
  pdfauthor={},
]{hyperref}
\usetikzlibrary{quantikz2}
\usepackage{authblk}

\newtheorem{lemma}{Lemma}

\newtheorem{remark}{Remark}

\title{Fourier space readout method for efficiently recovering functions encoded in quantum states}

\author[1,2]{Xinchi Huang\footnote{Email: kkou@quemix.com; huangxc@g.ecc.u-tokyo.ac.jp}}
\author[1,2]{Hirofumi Nishi}
\author[1,2]{Yoshifumi Kawada}
\author[3]{Tomofumi Zushi}
\author[1,2,4,5]{Yu-ichiro Matsushita}
\affil[1]{Department of Physics, The University of Tokyo, Tokyo 113-0033, Japan}
\affil[2]{Quemix Inc., Taiyo Life Nihombashi Building, 2-11-2 Nihombashi Chuo-ku, Tokyo 103-0027, Japan}
\affil[3]{Sumitomo Rubber Industries, Ltd., 2-1-1 Tsutsui, Chuo, Kobe, Hyogo 651-0071, Japan}
\affil[4]{Quantum Materials and Applications Research Center, National Institutes for Quantum Science and Technology (QST), 2-12-1 Ookayama, Meguro-ku, Tokyo 152-8550, Japan}
\affil[5]{Laboratory for Materials and Structures, Institute of Innovative Research, Tokyo Institute of Technology, Yokohama 226-8503, Japan}

\date{July 2025}

\begin{document}

\maketitle
\begin{abstract}
Applying quantum computing in the computer-aided engineering (CAE) problems are highly expected since quantum computers yield potential exponential speedups for the operations between extremely large matrices and vectors. Although efficient quantum algorithms for the above problems have been intensively investigated, it remains a crucial task to extract all the grid-point values encoded in the prepared quantum states, which was believed to eliminate the achieved quantum advantage. 
In this paper, we propose a quantum-classical hybrid Fourier space readout (FSR) method to efficiently recover the underlying function from its corresponding quantum state. We provide explicit quantum circuits, followed by theoretical and numerical discussions on its complexity. In particular, the complexity on quantum computers has only a logarithmic dependence on the grid number, while the complexity on classical computers has a linear dependence on the number of target points instead of the grid number. Our result implies that the achieved quantum speedups are not necessarily ruined when we read out the solutions to the CAE problems. 
\end{abstract}

\section{Introduction}

In the last decade, quantum computing has gathered increasing interest as the quantum computers are expected to provide potential exponential speedups for many classically time-consuming problems \cite{Childs.2003,Yuan2020,Babbush.2023,Lee.2023}. 
Compared to classical computers, the essential power of the quantum computers is the memory saving and acceleration of the matrix operations for extremely large matrices. For example, an $N\times N$ matrix can be interpreted as the quantum gate operation on only $\log_2 N$ qubits, possibly with several ancillary qubits if the matrix is non-unitary. Due to this fact, beyond the quantum chemistry problems \cite{Yuan2020,Lee.2023,Kassal.2008,Jones.2012,Ollitrault.2020,Childs.2022,Kosugi.2022,Kosugi.2023,Nishi.2023,Chan.2023,Brinet.2024, Huang.2024}, quantum computing is also highly expected in the computer-aided engineering (CAE) \cite{Kadowaki2025pre} in which faster and/or more accurate solving of the linear systems deduced from (possibly nonlinear) partial differential equations (PDEs) is one of the crucial tasks. 

The Harrow-Hassidim-Lloyd (HHL) algorithm \cite{Harrow.2009} gave a pioneer work for solving linear systems on quantum computers, which exploited the potential exponential speedup regarding the size of the matrix (the grid number in many applications). Since then, there are numerous works, known as the quantum linear system algorithms (QLSAs), to improve the HHL algorithm \cite{Ambainis2012,Clader.2013,Berry2014,Berry.2017,Childs.2017,Kieferova.2019,Childs.2020,Lin.2020,Childs.2021,Liu.2021,An.2022,Costa.2022,Fang.2023,Krovi2023,Berry.2024}, and some of them also addressed the nonlinear differential equations. 
Due to the linear dependence on the condition number, the best QLSAs (e.g., \cite{Costa.2022,Krovi2023}) have a polynomial speedup compared to the conventional classical algorithm by the sparse conjugate gradient (CG) method \cite{Shewchuk1994}. 
On the other hand, there were other algorithms specified for time evolution equations by a specific Schr\"odingerisation method \cite{Jin.2023} or a probabilistic imaginary-time evolution (PITE) method \cite{Huang.2024pre}, which achieved the exponential speedup in the grid number at the cost of a worse dependence on the error bound. Many other works \cite{Kiani.2022,An.2022pre,An.2023,Bagherimehrab.2023pre,Sato.2024,Sato.2024pre,Jin.2024,Sanavio.2024,Li.2025,Bharadwaj.2025,Wu.2025pre} also addressed the quantum speedups in preparing the quantum states corresponding to the solutions to PDEs. 
The above-mentioned quantum speedups cover the cost only for the quantum solving part including the query complexity of the oracles that encode the input vectors (initial-boundary conditions or the non-homogeneous terms) and the matrices (derived by discretizations). 
Although the precise preparation of such vector oracles and matrix oracles has the complexity $O(N)$ and $O(N^2)$, respectively \cite{Mottonen.2004}, the gate complexity of such oracles can be reduced to $O(\mathrm{polylog}N)$ by various approximation techniques in practical settings: \cite{Mozafari.2022,Nakaji.2022,Ramos2022,Moosa.2023,Bharadwaj.2023,Kosugi.2024,Daimon.2024,Conde.2024,Zylberman.2024} for the vector oracles and \cite{Camps.2024,Sunderhauf.2024,Guseynov.2024pre} for the matrix oracles. 
This implies that using good quantum algorithms, we can prepare the quantum states corresponding to the solutions at least polynomially faster (regarding the grid number $N$) than the conventional classical algorithms, e.g., the sparse CG method. 

Whereas, it remains a crucial issue to extract the classical information of the solutions from the quantum state, which is known as the readout problem. Whatever the quantum amplitude estimation \cite{Brassard.2002,Manzano.2023,Maronese.2024} or quantum state tomography \cite{Vogel.1989} techniques are applied, the quantum complexity of reading out the quantum state to obtain the values of solution at grid points has a polynomial scaling in $N$. This eliminates the achieved quantum advantage in the quantum solving part unless the solution is localized with sparse nonzero amplitudes \cite{Chen.2024}. 
Most of the previous works compromise to restrict the quantum advantage to the problems when only a finite number of observables (e.g., mean values of physical quantities) are necessary. 
However, the solutions themselves at the grid points are of great interest for a large amount of practical problems in computational fluid dynamics (CFD)/CAE simulations. 

To address the readout problem, a recent interesting direction is the idea of basis function expansion (a kind of feature extraction). For example, in \cite{Nishi.2025pre}, the solution (or the target state) is assumed to be approximately expressed by a linear combination of finite number of basis functions such as the Lorentzian functions (LFs). Utilizing the efficient encoding techniques for the LFs \cite{Kosugi.2024,Kosugi.2025}, the authors calculated the quantum overlaps between the target state and the states for the LFs, and solved relatively small classical optimization problems based on the derived overlaps to determine the weight, the shape and the peak of each LF. As long as the above parameters, understood as features, are determined, the approximate solution is known, and one can recover the values at any target grid points on classical computers. 

Besides, a similar idea was discussed in \cite{Miyamoto.2023} where the authors avoided the assumption and used some orthogonal function expansion to approximate the solution. The advantage of the expansion using orthogonal/orthonormal functions such as the trigonometric functions and the Chebyshev polynomials is that they form a complete orthogonal basis, so that the weight of each basis function can be obtained by simply calculating the overlap between the solution and the state for the basis function without solving optimization problems classically. However, for each basis, one needs to estimate the quantum overlap, and hence, it is a large cost especially for large dimensions unless there are only limited weights and their corresponding indices are known in advance. To avoid this problem, \cite{Miyamoto.2023} proposed a parameterized quantum circuit that was implemented with the help of tensor networks. By calculating a couple of suitable observables on quantum computers, the approximate weights encoded in the tensor network can be obtained by solving optimization problems on classical computers iteratively.  

The above idea of basis function expansion encounters the following difficulties in practice:  

\noindent (a) Whether we have the a priori knowledge of an efficient (small number of effective weights) linear combination of basis functions.  

\noindent (b) Whether the quantum circuits for the basis functions are quantum efficient (that is, the gate complexity is smaller than $O(\mathrm{polylog}N)$).

\noindent (c) How to efficiently determine the weights or parameters and estimate the quantum and/or classical complexity regarding both the grid number and the error bound.  

The proposal in \cite{Miyamoto.2023} addressed (a) and does not need any assumptions of the solution due to the choice of complete orthogonal functions. On the other hand, \cite{Nishi.2025pre} addressed (b) since the (discrete) LFs can be precisely and efficiently implemented on quantum computers. Both works used the quantum computers to obtain suitable quantum overlaps and solved optimization problems iteratively on classical computers. Roughly speaking, for one-dimensional case, the quantum complexity is $O(M^2 N_{\text{iter}} N_{\text{shot}})$, and the classical complexity is $O(M^3 N_{\text{iter}})$ where $M$ is the number of effective weights, $N_{\text{iter}}$ is the number of iterations in the optimization, and $N_{\text{shot}}$ is the number of repetitions for the quantum circuits. 

In this paper, we propose a novel method that simultaneously addresses all the three issues. We use the Fourier series expansion, which is complete, to approximate the solution so that we do not need any a priori knowledge of the efficient function bases. 
Thanks to the well-known efficient quantum circuit of the quantum Fourier transform (QFT), we overcome (b). Since QFT is a unitary, we can easily construct its inverse operator. Whereas, for other orthogonal functions such as the Chebyshev polynomials, the Legendre polynomials, and the orthogonal wavelets, the property of non-unitary yields further difficulties in constructing its inverse, and this requires more complicated discussion on the success probability of such a block encoding. 
For the issue (c), we adopt a completely different strategy without any classical optimizations. Different from the above-mentioned overlap-based methods that extract the data 
one by one, we obtain all the necessary Fourier coefficients (weights for the function bases) at the same time by repeating measurements of only one/two quantum circuit(s). This is similar to a direct way to read out the values at all grid points. For a large grid number $N$, the direct readout method (see Sect.~\ref{subsec:2-1}) suffers from the linear dependence on $N$. On the other hand, the Fourier coefficients are localized for continuous functions, and we show numerically in Sect.~\ref{sec3} and theoretically in Appendix \ref{sec:appC} that the quantum/classical complexity has no dependence on the grid parameter. 
This is a crucial message for the practical applications in the CFD/CAE simulations where the knowledge of the solutions themselves is required. Our result implies that similar to the state preparation problem of a large function-based vector, the cost for the readout problem depends only on the underlying function and the desired error bound, but is independent of the grid number. Therefore, the readout of the function-based quantum states does not ruin the quantum speedup regarding the grid number in the quantum solving part. 

The paper is organized as follows. 
In Sect.~\ref{sec2}, we propose the Fourier space readout (FSR) method, which is more efficient than the conventional method for deriving the quantum state that is governed by an underlying continuous function. Moreover, we introduce an additional extension operator to make the Fourier coefficients real-valued so that we can provide explicit and simple quantum circuits for deriving the Fourier coefficients. 
Section \ref{sec3} is devoted to the complexity of the proposed method, in which the quantum costs are estimated numerically for several examples. 
Together with the numerical reconstructions for 1D/2D functions in Sect.~\ref{sec3.5}, we demonstrate the great performance of our proposal, especially for large grid numbers. 
In Sect.~\ref{sec4}, we discuss some variations of the proposed method, including an overlap-based fully quantum Fourier space readout and an adaptive algorithm for the FSR method. The proposed adaptive algorithm in Sect.~\ref{subsec:4-2} shows how to determine the approximation parameter without the a priori information of the function and is hence, extremely useful in practice.  
We summarize the paper in Sect.~\ref{sec5} as the conclusion and provide theoretical details, as well as some numerical examples, in Appendices \ref{sec:appA}--\ref{sec:appE}.  

\section{Methods}
\label{sec2}

In this section, we introduce the method of the Fourier space readout. The essential idea is simple as the Fourier transform of a continuous function will be localized in the Fourier domain, which is easier to obtain. 
Moreover, we propose a detailed quantum circuit for the readout of the quantum states that are given by some underlying real-valued functions.   

\subsection{Real space readout}
\label{subsec:2-1}

Let the grid number be $N=2^n$ with the number of qubits $n\in \mathbb{N}$. Assume that we are given the following input quantum state:
$$
\ket{\psi}_n = \sum_{j=0}^{N-1} \psi_j \ket{j}_n.
$$
The aim is to obtain the approximations of the coefficients $\{\psi_j\}$ at all the grid points $j\in \{0,\ldots,N-1\}$ or some target grid points $j\in \mathcal{J}\subset \{0,\ldots,N-1\}$. 

\begin{figure}
\centering
\begin{quantikz}
\lstick[4]{\ket{\psi}} & \meter{0/1} & \rstick[4]{$|\psi_{0}|,\ldots,|\psi_{N-1}|$} \\[-0cm]
 & \meter{0/1} & \\[-0cm]
\setwiretype{n} & \vdots & \\
 & \meter{0/1} & 
\end{quantikz}
\caption{Quantum circuit for the real space readout.}
\label{sec2:fig1}
\end{figure}
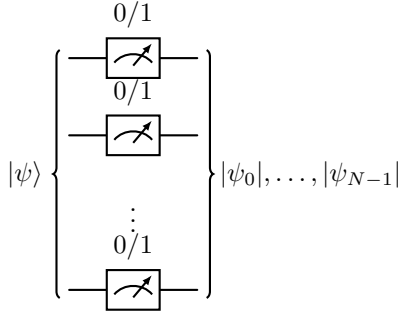

A conventional and direct way is to execute $Z$-basis measurements for all the qubits, and then by the repetitions of measurements, we can construct a histogram to obtain the absolute square of each amplitude, that is, $|\psi_j|^2$ for $j=0,\ldots,N-1$, if only the absolute values are of interest. Alternatively, we can do multiple positive operator-valued measures (POVMs) to obtain also the angles of the complex numbers $\psi_j$ by the change of basis. Considering the application such as the readout of a real-valued function at grid points, we call the above fundamental method the real space readout (RSR) in this paper, see Fig.~\ref{sec2:fig1}.  

\subsection{Fourier space readout (FSR)}
\label{subsec:2-2}

The RSR method is not efficient for the case of large $N$ because the number of repetitions needs to be $O(N)$ in general. 
In this paper, we consider the quantum state that is based on an unknown (continuous) real-valued function $f: [0,L]\rightarrow \mathbb{R}$ where $L>0$ is the size of the domain. In other words, we let 
$$
\psi_j = \frac{f(x_j)}{A_N}, \quad x_j = jL/N, j=0,\ldots,N-1,
$$
where $A_N := \left(\sum_{j=0}^{N-1}|f(x_j)|^2\right)^{1/2}$ is the normalized factor. Note that due to the principle of the quantum computers, the normalized factor $A_N$ is not distinguished and should be given as the a priori information (e.g., from the success probability if the input state is prepared by a quantum algorithm for solving the PDEs). 
For simplicity, we consider the one-dimensional case here, and the multi-dimensional cases can be similarly discussed, see Appendix \ref{sec:appA}. 

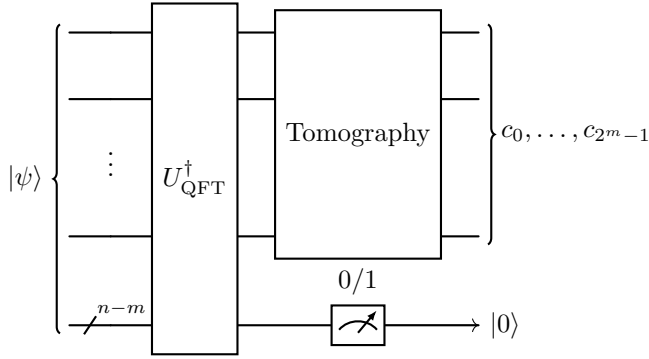
\begin{figure}
\centering
\begin{quantikz}
\lstick[5]{\ket{\psi}} &  & \gate[5]{U_{\text{QFT}}^\dag} & \gate[4]{\text{Tomography}} & \rstick[4]{$c_{0},\ldots,c_{2^m-1}$} \\[-0cm]
 &  &  &  &  \\[-0cm]
\setwiretype{n} & \vdots & \vdots &  & \\
 &  &  &  &  \\[-0cm]
 & \qwbundle{n-m} &  & \meter{0/1}\arrow[r] &
\rstick{\ket{0}} 
\end{quantikz}
\caption{Quantum circuit for the Fourier space readout with an integer parameter $m\le n$. We can apply the quantum tomography to obtain the first $2^m$ coefficients $c_0,\ldots,c_{2^m-1}$ by post-selecting the most significant $n-m$ qubits to be in $\ket{0}_{n-m}$ state.}
\label{sec2:fig2}
\end{figure}

We propose a Fourier space readout method, which is demonstrated in Fig.~\ref{sec2:fig2}. 
Here, $U_{\text{QFT}}^\dag$ is the inverse quantum Fourier transform (QFT), and we choose an integer $m\le n$, so that the measured information is compressed. The key idea of the FSR method is the Fourier series expansion of the function:
$$
f(x) = \sum_{k=-\infty}^\infty c_k \mathrm{exp}\left(\mathrm{i}\frac{2\pi}{L}kx\right). 
$$
According to the rapid decay of the Fourier coefficients $c_k$ (see Appendix \ref{subsec:C-1}), we can reconstruct the function $f$ approximately by 
\begin{equation}
\label{sec2:eq-cfr}
f(x) \approx \sum_{k=-(M-1)}^{M-1} c_k \mathrm{exp}\left(\mathrm{i}\frac{2\pi}{L}kx\right) = c_0 + 2\sum_{k=1}^{M-1} \mathcal{R}\left(c_k\mathrm{exp}\left(\mathrm{i}\frac{2\pi}{L}kx\right)\right), \quad x\in [0,L],
\end{equation}
where $M=2^m$ and $\mathcal{R}$ denotes the real part of a complex number. 
In this way, we can calculate the values at any grid point (even at finer grid points) by a parallel computation on classical computers. The classical computational cost scales as $O(JM)$, and the computational time is $O(M)$, where $J$ denotes the the number of target points that we are interested in. 
On the other hand, for the quantum computing part, the repetition number scales as $O(M)$, which is independent of the grid number $N$. Therefore, the FSR method is more efficient and more flexible compared to the conventional RSR method. The choice of the parameter $m$ and the detailed complexity will be provided in Sect.~\ref{sec3}. 

\subsection{Explicit quantum circuits for the FSR method}
\label{subsec:2-3}

In Fig.~\ref{sec2:fig2}, the dominant Fourier coefficients can be obtained by multiple POVMs in general. However, it seems not easy to accurately determine the angles of the complex numbers as we need to consider the stochastic error (i.e., sampling error) from the measurement. 
Under our setting that $f$ is a real-valued function, we propose the explicit quantum circuits for the FSR method. 

\begin{figure}
\centering
\begin{quantikz}
\lstick[4]{$\ket{0}_n$} &  \gate[5]{U_{\text{ext},\psi}} & \\
&  & \\
\setwiretype{n} & \vdots &  \\
&  & \\
\lstick{\ket{0}} & \qw &  
\end{quantikz}
= 
\begin{quantikz}
&  & \gate[4]{U_\psi} & \targ{} & \gate[4]{U_{+1}} & \\[-0cm]
&  &  & \targ{} &  & \\[-0cm]
\setwiretype{n} & \vdots &  & \vdots &  & \\
&  &  & \targ{} &  & \\[-0cm]
& \qw & \gate{H} & \ctrl{-4} & \ctrl{-1} & 
\end{quantikz}
\caption{A quantum circuit for the (even) extension operator $U_{\text{ext},\psi}$. Here, $U_{+1}$ denotes the quantum incrementer gate. }
\label{sec2:fig3}
\end{figure}
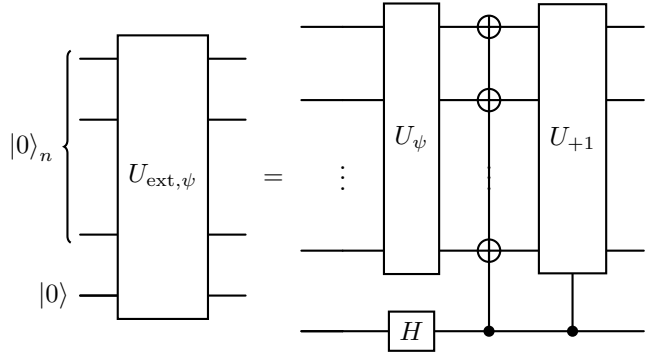

We assume that the oracle of preparing the input quantum state is given by $U_\psi$: 
$$
U_{\psi} \ket{0}_n := \ket{\psi}_n.
$$
We employ one ancillary qubit and introduce the following (even) extension operator:
\begin{equation}
\label{sec2:eq-ext}
U_{\text{ext},\psi} \left(\ket{0} \otimes \ket{0}_n\right) = \frac{1}{\sqrt{2}A_N}\left(\sum_{j=0}^{N-1} f(x_j)\ket{j}_{n+1} + \sum_{j=0}^{N-1} f(x_{N-j})\ket{N+j}_{n+1}\right).
\end{equation}
Assuming $f(x_N)=f(x_0)$, the above extension operator can be realized by the quantum circuit in Fig.~\ref{sec2:fig3}.
Here, we used a quantum fan-out gate (a series of CNOT gates) \cite{Kosugi.2024,Takahashi.2021}, as well as a controlled quantum incrementer gate \cite{Gidney2015pre,Yuan.2023} where the quantum incrementer gate $U_{+1}$ is defined by $U_{+1}\ket{k}_n := \ket{k+1\ \mathrm{mod}\ N}_n$. 
Using the extension operator, the extended quantum state becomes 
$$
\ket{\tilde \psi}_{n+1} = \frac{1}{\sqrt{2}A_N}\sum_{j=0}^{2N-1} \tilde{f}(\tilde x_j) \ket{j}_{n+1}, \quad \tilde x_j = jL/(2N), j=0,\ldots,2N-1,
$$
where $\tilde f$ is the even extension of $f$: $\tilde f(x) = f(x)$ for $x\in [0,L]$, and $\tilde f(x) = f(2L-x)$ for $x\in [L,2L]$. 
Thanks to such a symmetric structure, we derive real-valued Fourier coefficients after applying the inverse QFT. The verification is provided in Appendix \ref{sec:appB} for the completeness. 
Besides, we can obtain the quantum circuit for an odd extension (see Appendix \ref{sec:appB}) in a similar way. The odd extension yields better regularity across the boundary in the case that the boundary values are zero and the boundary derivatives are opposite (e.g., convex center-symmetric quadratic functions). 
\begin{remark}[Boundary conditions for CAE simulations]
The quantum circuit in Fig.~\ref{sec2:fig3} realizes the even extension in Eq.~\eqref{sec2:eq-ext} provided that $f(L)=f(0)$, which is automatically satisfied in the cases of periodic boundary conditions or homogeneous Dirichlet boundary conditions. 
Recalling that the input quantum state does not possess the value at the right boundary $x=L$. For more complicated boundary conditions, e.g., non-homogeneous Dirichlet conditions, one needs an additional multi-controlled single-qubit gate to input the value at the boundary $x=L$. An alternative way to deal with general boundary conditions is to use one ancillary qubit to evenly/oddly extend the solution to the underlying PDE in the quantum solving part so that we no longer need the extension operator in the readout part. 
\end{remark}
Next, we provide two quantum circuits to determine the absolute values and the signs of the dominant Fourier coefficients $\tilde c_k$, $k=0,\ldots, M-1$. 

\noindent \underline{Determination of the absolute values}

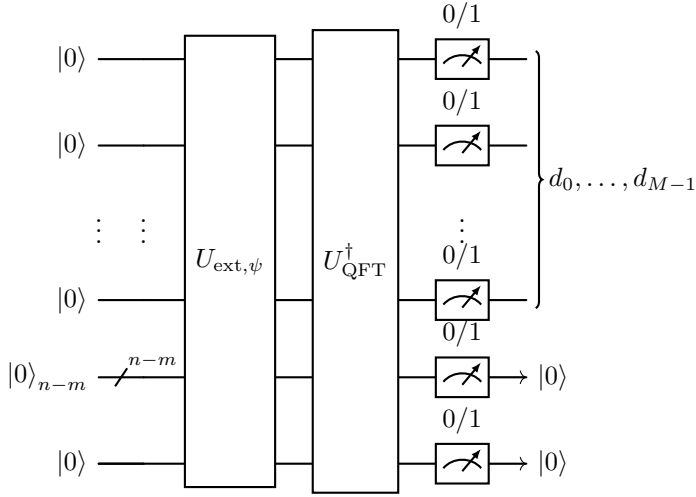
\begin{figure}
\centering
\begin{quantikz}
\lstick{\ket{0}} &  & \gate[6]{U_{\text{ext},\psi}} & \gate[6]{U_{\text{QFT}}^\dag} & \meter{0/1} & \rstick[4]{$d_{0},\ldots,d_{M-1}$} \\[-0cm]
\lstick{\ket{0}} &  &  &  & \meter{0/1} &  \\[-0cm]
\setwiretype{n}\vdots & \vdots &  & \vdots & \vdots & \\
\lstick{\ket{0}} &  &  &  & \meter{0/1} &  \\[-0cm]
\lstick{$\ket{0}_{n-m}$} & \qwbundle{n-m} &  &  & \meter{0/1}\arrow[r] &
\rstick{\ket{0}} \\
\lstick{\ket{0}} & \qw &  &  & \meter{0/1}\arrow[r] &
\rstick{\ket{0}} 
\end{quantikz}
\caption{Quantum circuit for determining the absolute values of the Fourier coefficients in the FSR method. By post-selecting the last $n-m+1$ qubits as $\ket{0}_{n-m+1}$, we repeat the measurements and obtain the absolute values of the Fourier coefficients $d_j \approx |\tilde c_j|$ for $j=0,\ldots,M-1$.}
\label{sec2:fig4}
\end{figure}

For the absolute values, we apply the inverse QFT and execute $Z$-basis measurements for the $n+1$ qubits as shown in Fig.~\ref{sec2:fig4}. 
By repeating the quantum circuit $N_{\text{shot1}}$ times and post-selecting the last (i.e., most significant) $n-m+1$ qubits to be in the state $\ket{0}_{n-m+1}$, we derive a real-valued vector $\mathbf{d} = \mathbf{d}(N_{\text{shot1}}) = (d_0,\ldots, d_{M-1})$ where $d_j$ approximates the absolute values of the Fourier coefficients $|\tilde c_j|$ for all $j=0,\ldots,M-1$. 

\noindent \underline{Determination of the signs}

\begin{figure}
\centering
\resizebox{15cm}{!}{
\begin{quantikz}
\lstick{\ket{0}} &  & \gate[6]{U_{\text{ext},\psi}} & \gate[6]{U_{\text{QFT}}^\dag} & \gate[4]{H^{\otimes m}} &  & \meter{0/1} & \rstick[4]{$e_{0},\ldots,e_{M-1}$} \\[-0cm]
\lstick{\ket{0}} &  &  &  &  &  & \meter{0/1} &  \\[-0cm]
\setwiretype{n}\vdots & \vdots &  & \vdots &  & \vdots & \vdots & \\
\lstick{\ket{0}} &  &  &  &  &  & \meter{0/1} &  \\[-0cm]
\lstick{\ket{0}} & \qwbundle{n-m} &  &  &  &  & \meter{0/1}\arrow[r] &
\rstick{\ket{0}} \\
\lstick{\ket{0}} & \qw &  &  &  &  & \meter{0/1}\arrow[r] &
\rstick{\ket{0}} \\
\lstick{\ket{0}} & \gate{H} & \ctrl{-1} & \ctrl{-1} & \octrl{-3} & \gate{H} & \meter{0/1}\arrow[r] &
\rstick{\ket{0}} 
\end{quantikz}
$\equiv$
\begin{quantikz}
\lstick{\ket{0}} &  & \gate[6]{U_{\text{ext},\psi}} & \gate[6]{U_{\text{QFT}}^\dag} &  &  & \meter{0/1} & \rstick[4]{$e_{0},\ldots,e_{M-1}$} \\[-0cm]
\lstick{\ket{0}} &  &  &  &  &  & \meter{0/1} &  \\[-0cm]
\setwiretype{n}\vdots & \vdots &  & \vdots & \vdots & \vdots & \vdots & \\
\lstick{\ket{0}} &  &  &  &  &  & \meter{0/1} &  \\[-0cm]
\lstick{\ket{0}} & \qwbundle{n-m} &  &  & \gate[2]{H^{\otimes (n-m+1)}} &  & \meter{0/1}\arrow[r] &
\rstick{\ket{0}} \\
\lstick{\ket{0}} & \qw &  &  &  &  & \meter{0/1}\arrow[r] &
\rstick{\ket{0}} \\
\lstick{\ket{0}} & \gate{H} & \ctrl{-1} &   & \octrl{-1} & \gate{H} & \meter{0/1}\arrow[r] &
\rstick{\ket{0}} 
\end{quantikz}
}
\caption{Auxiliary quantum circuit for determining the signs of the dominant Fourier coefficients in the FSR method. The right quantum circuit is equivalent to the left one, and we use simply a QFT instead a controlled QFT. By post-selecting the last $n-m+2$ qubits as $\ket{0}_{n-m+2}$, we obtain $e_j \approx \frac{1}{2}|\tilde c_j + 1/\sqrt{M}|$ for $j=0,\ldots,M-1$.}
\label{sec2:fig5}
\end{figure}
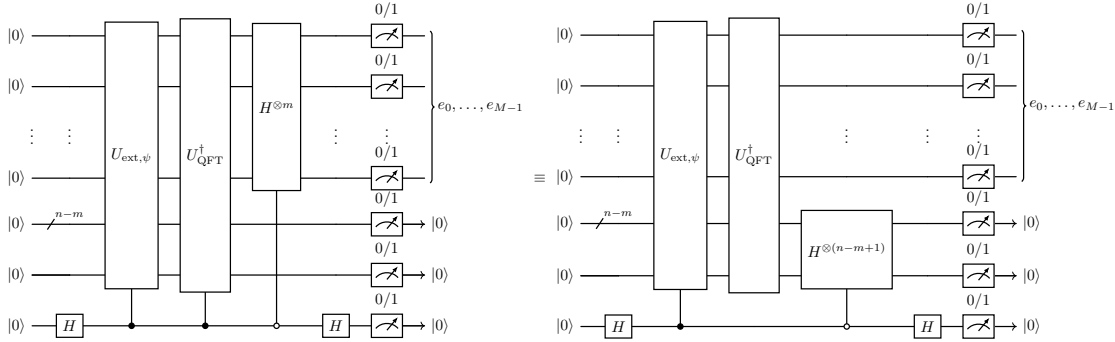

To determine the signs, we propose an auxiliary quantum circuit. The idea is to shift the quantum state by adding a constant for the dominant amplitudes. To this end, we consider a quantum circuit of linear combination of unitaries (LCU) \cite{Childs.2012}. The quantum circuit is shown in Fig.~\ref{sec2:fig5}. 
By repeating the quantum circuit $N_{\text{shot2}}$ times and post-selecting the last (i.e., most significant) $n-m+2$ qubits to be in the state $\ket{0}_{n-m+2}$, we derive a real-valued vector $\mathbf{e}^{(\ell)} = \mathbf{e}^{(\ell)}(N_{\text{shot2}}) = (e_0^{(\ell)},\ldots, e_{M-1}^{(\ell)})$ where $e_j^{(\ell)}$ approximates $\frac{1}{2}|\tilde c_j + 1/\sqrt{M}|$ for all $j=0,\ldots,M-1$. 
Here, we take $\ell=1,\ldots,N_{\text{iter}}$ with a given iteration number $N_{\text{iter}}$, and we define 
$$
\mathbf{e} := \frac{1}{N_{\text{iter}}}\sum_{\ell=1}^{N_{\text{iter}}} \mathbf{e}^{(\ell)}, 
$$
to reduce the stochastic error to some extent. Although we introduce the parameter $N_{\text{iter}}$ for generality, we simply take $N_{\text{iter}}=1$ throughout this paper. 
Moreover, we introduce a real-valued vector:
$$
\mathbf{g} = \mathbf{g}(N_{\text{shot1}}, N_{\text{shot2}}, N_{\text{iter}}) := 2\mathbf{e}-\mathbf{d}-\left(1/\sqrt{M},\ldots, 1/\sqrt{M} \right) = (g_0,\ldots, g_{M-1}). 
$$
For a given small $\delta>0$, we determine the sign of $\tilde c_k$ as follows:
\begin{itemize}
\item The sign of $\tilde c_k$ is set to plus if $g_k>-\delta$;

\item The sign of $\tilde c_k$ is set to minus if $g_k\le -\delta$.
\end{itemize}
If there are no sampling errors, that is, $N_{\text{shot1}}, N_{\text{shot2}}\to \infty$, then $\mathbf{d} = (|\tilde c_0|, \ldots, |\tilde c_{M-1}|)$ and $\mathbf{e} = \frac12 (|\tilde c_0+1/\sqrt{M}|,\ldots, |\tilde c_{M-1}+1/\sqrt{M}|)$. 
Therefore, we have
$$
g_k = 
\left\{
\begin{aligned}
& 0, &\quad \tilde c_k\ge 0, \\
& 2\tilde c_k<0, &\quad -1/\sqrt{M}<\tilde c_k< 0, \\
& -2/\sqrt{M}, &\quad \tilde c_k\le -1/\sqrt{M}.
\end{aligned}
\right.
$$
In the ideal cases without sampling errors, the coefficient $\tilde c_k$ is positive if $g_k = 0$, while it is negative if $g_k \le -2\min_{k}|\tilde c_k|< 0$. 
On the other hand, considering the sampling errors, we cannot determine exactly the signs of the coefficients whose absolute values are smaller than the inverse root of the number of repetitions. 
Fortunately, the sign inversions of such coefficients do not contribute much to the final reconstruction. The error can be controlled by taking large numbers of shots and choosing a suitable $\delta$. A practical choice of $\delta$ is of order $O(1/\sqrt{N_{\text{shot2}}})$, and we show its numerical performance in Sect.~\ref{sec3}. 
To sum up, we approximate the dominant Fourier coefficients by 
\begin{equation}
\tilde c_k \approx \mathrm{sgn}(g_k+\delta) d_k, \quad k=0,\ldots,M-1.
\end{equation}
Here, $\mathrm{sgn}(\cdot)$ is the sign function, but we practically set the sign to be minus instead of $0$ if $g_k=-\delta$. 
As long as we calculate the approximate Fourier coefficients, the real-valued function $f$ is recovered by the following formula:
\begin{equation}
\label{sec2:eq-rec}
f(x) = \tilde f(x) \approx \frac{A_N}{\sqrt{N}}\left(\tilde c_0 + 2\sum_{k=1}^{M-1} \mathcal{R}\left(\tilde c_k\mathrm{exp}\left(\mathrm{i}\frac{2\pi}{2L}kx\right)\right)\right) 
= \frac{A_N}{\sqrt{N}}\left(\tilde c_0 + 2\sum_{k=1}^{M-1}\tilde c_k \cos\left(\frac{k\pi}{L}x\right)\right), 
\end{equation}
for $x\in [0,L]$. Here, the formula is different from Eq.~\eqref{sec2:eq-cfr}, and we have a pre-factor $A_N/\sqrt{N}$. The reason is that the Fourier coefficients we obtained by the quantum computing have a constant scaling compared to the continuous Fourier coefficients (see the last part of Appendix \ref{sec:appB} and Eq.~\eqref{appB:eq-rec} therein for the details where $\tilde N=2N$, $\tilde L=2L$, $\tilde A_{\tilde N}=\sqrt{2}A_N$ due to the extension).  

\begin{remark}[Reduction of measured qubits]
If $M$ is sufficiently large such that the Fourier coefficients after the index $M/2$ are negligible, then we can approximately remove the $Z$-basis measurements for the registers of $n-m$ qubits in Figs.~\ref{sec2:fig4} and \ref{sec2:fig5}. Instead, we need an additional (controlled) shift gate $U_{\mathrm{shift}}(M/2):$ $U_{\mathrm{shift}}(M/2) \ket{k}_{n+1} := \ket{k+M/2}_{n+1}$ after the (controlled) inverse $\mathrm{QFT}$ to make all the dominant Fourier coefficients (including the duplicated ones) appear in the small indices.
\end{remark}
\begin{remark}[Utilization of other measurement results]
Assume that all the qubits are measured. In Fig.~\ref{sec2:fig5}, we post-select the last qubit to be $\ket{0}$. Indeed, we can obtain $\tilde e_j \approx \frac{1}{2}|\tilde c_j-1/\sqrt{M}|$ for $j=0,\ldots,M-1$ if the post-selection is $\ket{1}$. This information could be used to relieve the stochastic error in the determination of the signs. 
Moreover, we post-select the last $n-m$ qubits in the input register and the second last qubit to be $\ket{0}_{n-m+1}$ in Fig.~\ref{sec2:fig5}. In fact, by noting the relation $c_{k}=\overline{c_{2N-k}}=c_{2N-k}$ for $k=1,\ldots,N$ and that the shift $1/\sqrt{M}$ only applies to the first $M\le N$ Fourier coefficients, we can also estimate the absolute value $|c_j|=|c_{2N-j}|$ for $j=1,\ldots,M-1$ by collecting the results that the second last qubit is measured to be $\ket{1}$ and the input register is measured to be $\ket{N-j}_n$ for $j=1,\ldots,M-1$. Thus, together with a technical post-processing to determine $|c_0|$ on classical computers, we can omit the quantum circuit in Fig.~\ref{sec2:fig4}. 
\end{remark}

\section{Complexity}
\label{sec3}

In this section, we discuss the complexity of the proposed FSR method.  
The gate complexity of the quantum fan-out gate, the quantum incrementer gate, and the QFT are at most $O\left(n^2\right)$ with the circuit depth $O(n)$. Assume that the (controlled) preparation oracle $U_\psi$ has the gate complexity $O(\mathrm{polylog} N)$. According to the explicit quantum circuits in Figs.~\ref{sec2:fig4} and \ref{sec2:fig5}, the gate complexity in each circuit is $O(\mathrm{polylog} N)$, which is quantum efficient. 
To estimate the complexity for the quantum computing, we need to evaluate the number of repetitions of the quantum circuits. 
In the following context, we evaluate the parameter $M$ and the number of repetitions, respectively, regarding the error bound $\varepsilon$ and the grid number $N$.  

\subsection{Evaluation of approximation parameter}
\label{subsec:3-1}

The key idea in this paper is to use the Fourier series expansion of a real-valued function. Recalling that $M=2^m$ for some $m\le n$ is an approximation parameter corresponding to the truncation of the Fourier series. 
By the theoretical analysis for the truncation error, we obtain an overhead of the approximation parameter:
\begin{equation}
\label{sec3:eq-M}
M = O((1/\varepsilon)^{s}),
\end{equation}
for some $s>0$. In the case that the function $f$ is absolutely continuous which is fulfilled for many applications, we have $0<s\le 2/3$, that is, a sublinear dependence on $1/\varepsilon$, see Lemma \ref{appC:lem1} in Appendix \ref{subsec:C-1}. Moreover, according to Childs et al. \cite{Childs.2022}, we can obtain a greatly improved overhead $M=O(\mathrm{polylog}(1/\varepsilon))$ provided that the function is smooth and satisfies the periodic boundary conditions. Here, we demonstrate two numerical examples, and we postpone the theoretical details to Appendix \ref{sec:appC}. 
\begin{figure}
\centering
\resizebox{14cm}{!}{
\includegraphics[keepaspectratio]{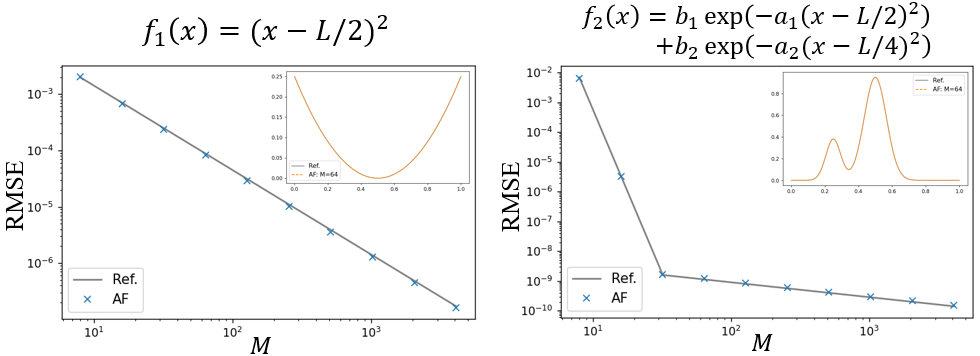}
}
\caption{RMSE plots regarding approximation parameter $M$ for two real-valued functions. The left subplot shows the case for a quadratic function where the gray reference line indicates $O(M^{-3/2})$. The right subplot shows the case for a linear combination of two Gaussian functions where the gray reference lines indicate $O(M^{-11})$ and $O(M^{-1/2})$, respectively. The blue cross marks imply the results using the approximate Fourier method (i.e., the FSR method without considering the sampling errors) with $M=8,16,32,\ldots,4096$. }
\label{sec3:fig1}
\end{figure}

The root mean square errors (RMSEs) for either a quadratic function or a linear combination of two Gaussian functions are shown in Fig.~\ref{sec3:fig1}. Here, we choose $L=1$, $a_1=1024/9$, $a_2=256$, $b_1=0.95$, $b_2=0.38$. The results for the FSR method (without considering the sampling errors) are obtained by the Qiskit simulator \cite{Qiskit2023} using its function \textbf{get\_statevector()}, so that only the approximation errors are addressed. For the quadratic function, the RMSE has the dependence $O(M^{-3/2})$, which implies $s\approx 0.67$ provided that the RMSE is upper bound by $\varepsilon$. Although the function is smooth, it has a discontinuous derivative at the endpoints if we regard it as a periodic function. This prevents us from a faster decay regarding $M$. 
For the second case of a linear combination of two Gaussian functions, the RMSE decreases rapidly for $M\le 32$ ($s\approx 0.09$) and decreases much slower for $M>32$ ($s\approx 2$). The fast decay could be understood as the nearly continuous derivatives across the boundary (i.e., the left derivative at the left endpoint is almost the same as the right derivative at the right endpoint). 
See Remark \ref{appC:rem2} in Appendix \ref{subsec:C-1} for the theoretical justifications. 

In the above examples, we find that a practical precision (e.g., $\varepsilon< 10^{-3}$) can be obtained by choosing relatively small $M$ (e.g., $M=64$). 

\subsection{Evaluation of number of repetitions}
\label{subsec:3-2}

In addition to the approximation error from the truncation of the Fourier series, there are also sampling errors when we retrieve the dominant Fourier coefficients from the corresponding quantum state. As we know, the repetitions of the measurements could relieve such stochastic errors. 

When a quantum state: 
$$
\ket{\psi} = \sum_{k=0}^{N-1} \psi_k \ket{k}_n 
$$
is prepared on a quantum computer, we repeat the $Z$-basis measurements of the $n$ qubits for $N_{\text{shot}}$ times. 
For each $k=0,\ldots,N-1$, we can regard the number of successes (obtaining the base $\ket{k}_n$) in a sequence of $N_{\text{shot}}$ experiments as a random variable $Z_k$. Note that the probability of success, obtaining the base $\ket{k}_n$, is $p_k = |\psi_k|^2$, and the probability of failure is $1-p_k$. Roughly speaking, the stochastic error for obtaining each $p_k$ is proportional to $\sqrt{p_k(1-p_k)/N_{\text{shot}}}$. 

In the case of the RSR method, the amplitude is the normalized grid-point value $\psi_k = f(x_k)/A_N$ for each $k=0,\ldots,N-1$, and hence, $p_k=|f(x_k)|^2/A_N^2$. Thus, the error for estimating $f(x_k)$ is multiplied by a factor $A_N$. To achieve an error bound $\varepsilon$, this requires $A_N^2\sqrt{p_k(1-p_k)/N_{\text{shot}}} = O(\varepsilon)$ and yields the order of the repetition number $N_{\text{shot}}=O(N(1/\varepsilon)^2)$. 
On the other hand, in the case of FSR method, the amplitude is the quantum Fourier coefficient $c_k$, which is a $N$-uniform scaling of the continuous Fourier coefficient (see Eq.~\eqref{appB:eq-rqc} in Appendix \ref{sec:appB}). Then, $p_k=|c_k|^2$ for each $k=0,\ldots,N-1$. According to the reconstruction formula \eqref{sec2:eq-rec}, there is no explicit dependence on $N$. Due to the approximation that we use only $M$ dominant Fourier coefficients, the dependence on the error bound $\varepsilon$ can be slightly worse, and the repetition number has the order: 
\begin{equation}
\label{sec3:eq-order}
N_{\text{shot}}=O((1/\varepsilon)^{2+\hat s}),
\end{equation}
for some $\hat s>0$, which is related to the factor $s$ in Eq.~\eqref{sec3:eq-M} ($\hat s=s$ theoretically). The detailed derivation of the overhead of $N_{\text{shot}}$ is provided in Appendix \ref{sec:appC}. 
Here, we demonstrate two examples to justify the numerical performance of the FSR method, addressed by the comparison to the RSR method. 

The first example is the quadratic function $f_1$, and the second one is the linear combination of two Gaussian functions as we discussed in the above subsection. We choose the RMSE between the exact vector $(f(x_1),\ldots,f(x_{N-1}))$ and the reconstructed vector using either the RSR method (Fig.~\ref{sec2:fig1}) or the FSR method using the quantum circuits in Figs.~\ref{sec2:fig4} and \ref{sec2:fig5}. For the FSR method, we choose $N_{\text{shot1}}=N_{\text{shot2}}=N_{\text{shot}}$, $N_{\text{iter}}=1$, and the desired vector is calculated by Eq.~\eqref{sec2:eq-rec} on classical computers after we obtain the Fourier coefficients via the quantum simulator Qiskit \cite{Qiskit2023}. 

\begin{figure}
\centering
\resizebox{14cm}{!}{
\includegraphics[keepaspectratio]{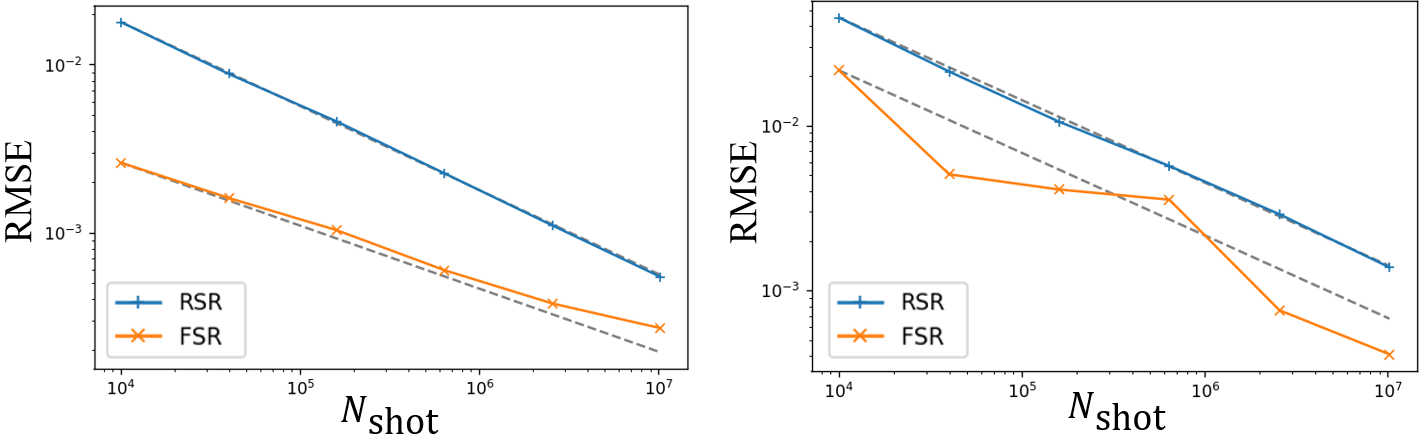}
}
\caption{RMSE plots regarding number of repetitions $N_{\text{shot}}$ for the RSR method and the FSR method. Two subplots for a quadratic function and a linear combination of two Gaussian functions are illustrated. In the left subplot, the gray reference lines indicate $O\left((N_{\text{shot}})^{-1/2}\right)$ and $O\left((N_{\text{shot}})^{-3/8}\right)$, respectively. In the right subplot, the gray reference lines indicate both $O\left((N_{\text{shot}})^{-1/2}\right)$.
The plus and cross marks imply the results for the RSR method and the approximate FSR method ($M=64$), respectively with $N_{\text{shot}}=10000,40000,160000,\ldots,10240000$.}
\label{sec3:fig2}
\end{figure}

First, we take $N=1024$, $M=64$ and discuss the dependence on the number of repetitions. The results are shown in Fig.~\ref{sec3:fig2}. 
In the case of the quadratic function, the RMSE scales as $O\left((N_{\text{shot}})^{-1/2}\right)$ for the RSR method, while it is almost $O\left((N_{\text{shot}})^{-3/8}\right)$ for the FSR method. The power $-3/8$ is deduced from the theoretical bound for the quadratic function. Here, we take a fixed number $M=64$, and the approximation error will be dominant for sufficiently large $N_{\text{shot}}$, which explains the last two points for the FSR method. To derive precisely the theoretical order, we need to choose adaptively $M=O\left((1/\varepsilon)^{2/3}\right)$. Although $M$ cannot be exactly determined for a given error bound, in Sect.~\ref{subsec:4-2}, we mention how to choose an adaptive $M$ corresponding to $N_{\text{shot}}$ in practice.   
In the case of the linear combination of two Gaussian functions, it seems that the RMSEs for both methods are proportional to $(N_{\text{shot}})^{-1/2}$. 
Figure \ref{sec3:fig2} verifies the dependence on the error bound (see Eq.~\eqref{sec3:eq-order}), and numerically $\hat s\approx 2/3$ for the quadratic function while $\hat s\approx 0$ for the linear combination of Gaussian functions. The better performance for the second function may come from the better approximation as we illustrated in Fig.~\ref{sec3:fig1} that the approximation error for the linear combination of Gaussian functions is already very small when we take $M=64$. 

\begin{figure}
\centering
\resizebox{14cm}{!}{
\includegraphics[keepaspectratio]{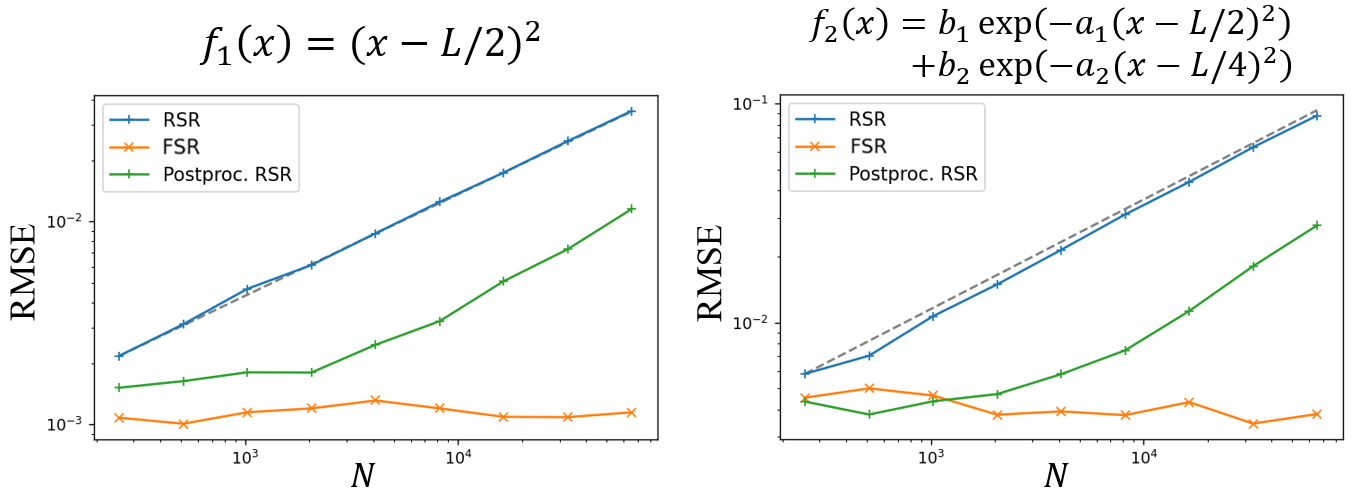}
}
\caption{RMSE plots regarding grid number $N$ for the RSR method and the FSR method. Two subplots for a quadratic function and a linear combination of two Gaussian functions are illustrated. The gray reference lines indicate $O\left(N^{1/2}\right)$. 
The plus and cross marks imply the results for the RSR method and the approximate FSR method ($M=64$), respectively with $N=256,512,1024,\ldots,65536$. Here, we also provide the results for the RSR method with a classical post-processing to reduce the stochastic errors. }
\label{sec3:fig3}
\end{figure}

Next, we take $N_{\text{shot}}=160000$, $M=64$ and discuss the scaling with respect to the grid number $N$. The results are shown in Fig.~\ref{sec3:fig3}.
In both cases, the RMSEs scale as $O\left(N^{1/2}\right)$ for the RSR method, while it has no dependence on $N$ for the FSR method.
Moreover, we give the results for a post-processed RSR method in green lines, where we apply the fast Fourier transform, cut off the high frequency components after $M=64$, and then apply the inverse fast Fourier transform to relieve the oscillations of the original RSR results. The post-processed RSR method seems good for small grid numbers, but it still has the order $O\left(N^{1/2}\right)$ for sufficiently large $N$. For the practical interest of applications, we provide the numerical comparison between the RSR method with the post-processing and the FSR method under a small repetition number $N_{\text{shot}}=1000$ in Appendix \ref{sec:appD}. 

To sum up, we find that numerically the RMSEs for the RSR method and the FSR method are $O\left(\sqrt{N}/\sqrt{N_{\text{shot}}}\right)$ and $O\left(1/N_{\text{shot}}^{s_0}\right)$ for some positive $s_0\le 1/2$, respectively. This verifies our assertions on the numbers of repetitions for both methods (see also Appendix \ref{subsec:C-2}). In Sect.~\ref{sec3.5}, we plot the numerical reconstructions for 1D/2D functions to show that the FSR method outperforms the RSR method, especially for large grid numbers.
\begin{remark}[Parameter choice in the determination of signs]
In Figs.~\ref{sec3:fig2} and \ref{sec3:fig3}, we use the explicit quantum circuits in Figs.~\ref{sec2:fig4} and \ref{sec2:fig5} for the FSR method, and choose $\delta = 2/\sqrt{N_{\mathrm{sum}}}$, where $N_{\mathrm{sum}}\le N_{\mathrm{shot}}$ counts the number of samples that the last $n-m+2$ qubits in Fig.~\ref{sec2:fig5} are post-selected to be $\ket{0}$. The signs could be inverse if the Fourier coefficients are smaller than the order $O(1/\sqrt{N_{\mathrm{shot}}})$, this partially leads to a relatively large variance compared to the RSR method. 
\end{remark}

\subsection{Computational resources for the FSR method}

We summarize the discussions in this section, and provide a comparison of the computational costs between the RSR method and the FSR method in Table \ref{sec3:tab1}.
\begin{table}[htb]
\centering
\caption{Classical and quantum resources for the RSR method and the FSR method. }
\label{sec3:tab1}
\scalebox{0.9}[0.9]{
\begin{tabular}{l|ccc}
\hline
Method & Ancillary count & Quantum complexity & Classical complexity \\
\hline
RSR & 0 & $\tilde{O}\left(N (1/\varepsilon)^{2}\right)$ & $O\left(N (1/\varepsilon)^{2}\right)$ \\
FSR & $2$ & $O\left(\mathrm{polylog} N (1/\varepsilon)^{2+s}\right)$ & $O\left((1/\varepsilon)^{2+s}\right) + O\left(J(1/\varepsilon)^{s}\right)$ \\
\hline
\end{tabular}
}
\end{table}
Here, we assume that the oracle for the input quantum state is prepared within the gate complexity $O(\mathrm{polylog}N)$. $N$ and $\varepsilon$ denote the grid number and the error bound, respectively. $J\le N$ is the number of target grid points, and $s$ is the positive constant related to the smoothness of the underlying function for the input quantum state, see Eq.~\eqref{sec3:eq-M}. As we mentioned, $s\le 2/3$ if the (1D) function is absolutely continuous and the values at both endpoints are equal. 
Details on the theoretical orders are provided in Appendix \ref{sec:appC}. For the $d$-dimensional case, the ancillary count should be changed into $d+1$ (see the quantum circuits for multi-dimensional FSR method in Appendix \ref{sec:appA}), and one can estimate the parameter $s$ similarly as the discussions for a one-dimensional function. 

The first term in the classical complexity corresponds to the cost of making a histogram from the measurements by the quantum computers, and hence, is proportional to the number of repetitions. The second term in the classical complexity for the FSR method comes from the derivation of the values at the target points whose cost is $O(MJ)$. The classical computation time can be further reduced to $O(M)$ if we execute the calculations in parallel.
Table \ref{sec3:tab1} shows that the FSR method greatly outperforms the RSR method (i.e., an exponential speedup) regarding the grid number at the cost of a slightly heavier dependence on the error bound. Moreover, since we extract the features of the underlying function (dominant Fourier coefficients), the proposed FSR method is more flexible and cheaper to obtain the values at $J$ target points as $J$ can be much smaller than $N$. This greatly meets the needs for some CAE simulations that only the solutions in the target domains are of interest. 


\section{Numerical examples}
\label{sec3.5}

To give a clear vision of our proposal, we provide the reconstructions of both 1D and 2D functions in this section. 
The numerical results are obtained using the quantum simulator Qiskit.
Moreover, to avoid the a priori information of the functions, we apply the adaptive algorithm in Sect.~\ref{subsec:4-2}, so that the approximation parameter $M$ is automatically determined according to the repetition number $N_{\text{shot}}$. 

\subsection{1D functions}

\begin{figure}
\centering
\resizebox{14cm}{!}{
\includegraphics[keepaspectratio]{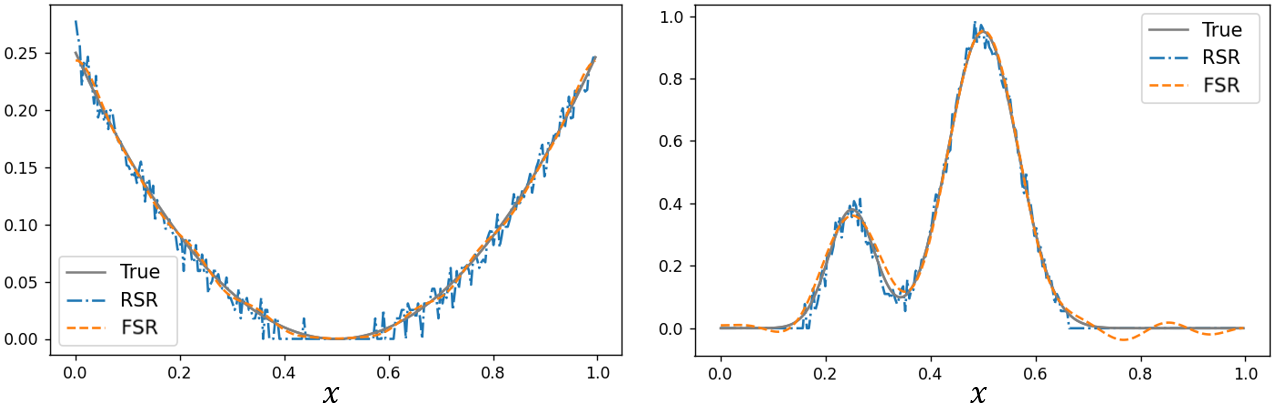}
}
\caption{Reconstructions by the RSR method and the FSR method for two functions with a small grid number $N=256$. The left subplot is for a quadratic function, while the right subplot is for a linear combination of Gaussian functions. Here, the number of repetitions is chosen as $N_{\text{shot}}=10^4$. }
\label{sec3.5:fig1}
\end{figure}

We consider the 1D functions discussed in the previous section. That is, 
\begin{align*}
& f_1(x) = (x-0.5)^2, \quad x\in [0,1],\\
& f_2(x) = 0.95\,\mathrm{exp}\left(-\frac{1024}{9}(x-0.5)^2\right) + 0.38\,\mathrm{exp}\left(-256(x-0.25)^2\right), \quad x\in [0,1].
\end{align*}
In Fig.~\ref{sec3.5:fig1}, we compare the FSR reconstructions with the RSR reconstructions under the choices of $N_{\text{shot}}=10^4$ and a small grid number $N=256$. Here, the adaptive approximation parameter is $M=32$ for both functions. 
For the quadratic function, the RMSE between the RSR reconstruction and the true solution is $8.934\times 10^{-3}$, while the RMSE between the FSR reconstruction and the true solution is $2.645\times 10^{-3}$. 
For the linear combination of Gaussian functions, the RMSE between the RSR reconstruction and the true solution is $2.145\times 10^{-2}$, while the RMSE between the FSR reconstruction and the true solution is $1.633\times 10^{-2}$. 
The plots in Fig.~\ref{sec3.5:fig1} show that the FSR method has a similar performance to the RSR method for a small grid number, while the FSR reconstructions are much more smooth. 

\begin{figure}
\centering
\resizebox{14cm}{!}{
\includegraphics[keepaspectratio]{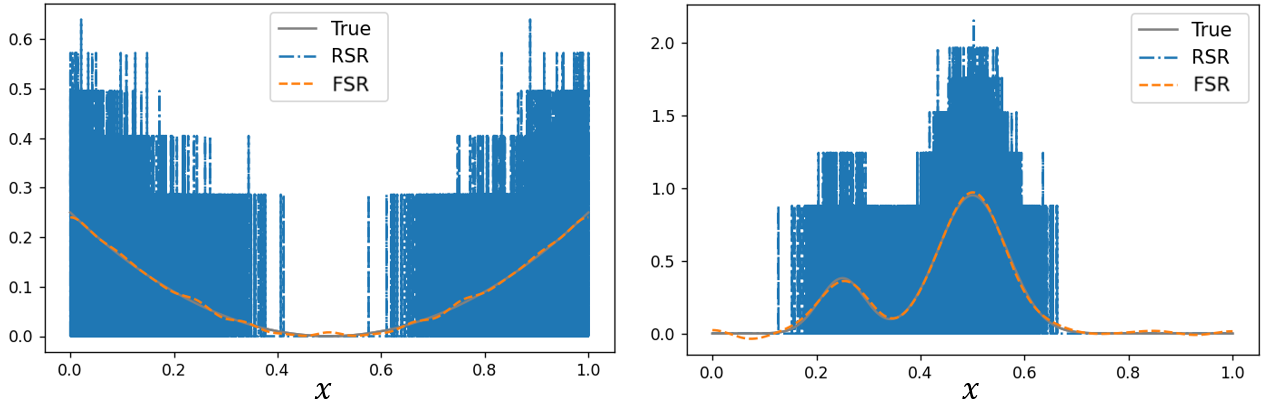}
}
\caption{Reconstructions by the RSR method and the FSR method for two functions with a large grid number $N=65536$. The left subplot is for a quadratic function, while the right subplot is for a linear combination of Gaussian functions. Here, the number of repetitions is chosen as $N_{\text{shot}}=10^4$. }
\label{sec3.5:fig2}
\end{figure}

On the other hand, we illustrate the reconstruction results under the choices of a large grid number $N=65536$ in Fig.~\ref{sec3.5:fig2}. The adaptive approximation parameter is automatically determined to be $M=32$ for both functions. For the FSR method, the RMSEs are $3.319\times 10^{-3}$ and $1.444\times 10^{-2}$, respectively, which are similar to those in the case of a small grid number $N=256$. However, the RSR reconstructions are highly oscillating and no longer make sense since the errors for the RSR method are proportional to $\sqrt{N}$. This implies the great performance of the proposed FSR method compared to the conventional one (RSR) for large grid numbers. 

\subsection{2D functions}

\begin{figure}
\centering
\resizebox{14cm}{!}{
\includegraphics[keepaspectratio]{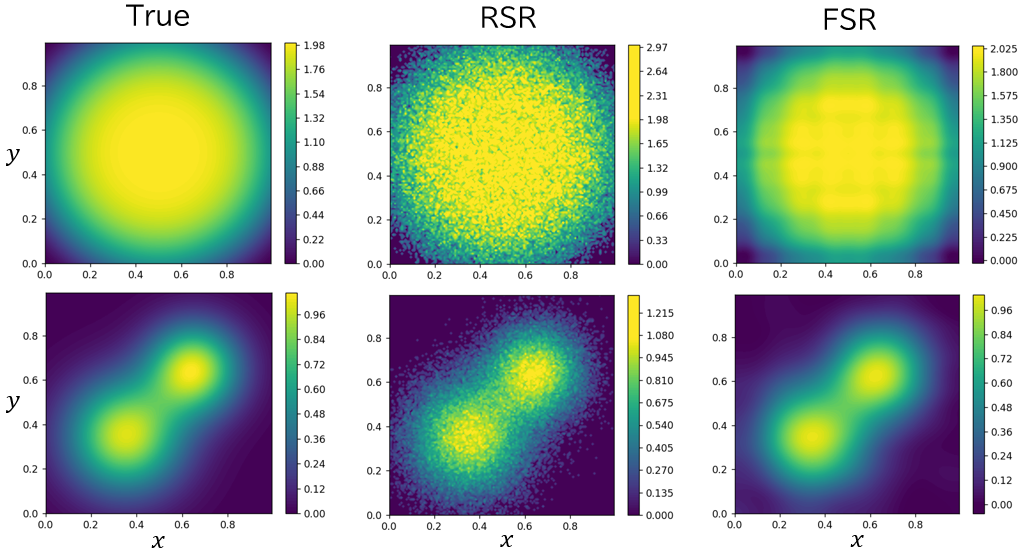}
}
\caption{Reconstructions by the RSR method and the FSR method for a 2D trigonometric function (upper subplots) and a linear combination of 2D Gaussian functions (lower subplots). Here, the grid number is $N=N_1\times N_2=128\times 128$, and the number of repetitions is chosen as $N_{\text{shot}}=10^5$. }
\label{sec3.5:fig3}
\end{figure}

\begin{figure}
\centering
\resizebox{14cm}{!}{
\includegraphics[keepaspectratio]{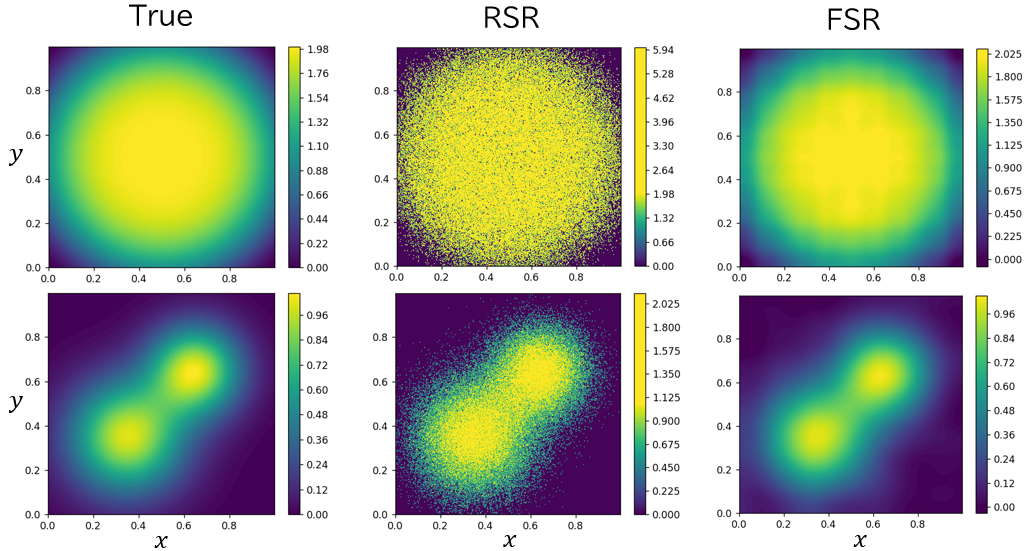}
}
\caption{Reconstructions by the RSR method and the FSR method for two functions with a large grid number $N=N_1\times N_2 = 512\times 512$. Here, the number of repetitions is chosen as $N_{\text{shot}}=10^5$. }
\label{sec3.5:fig4}
\end{figure}

We also provide two-dimensional numerical results for some specific functions in Fig.~\ref{sec3.5:fig3}. Here, the FSR method indicates the adaptive algorithm with a parameter $N_{\text{mar}}=4$ (see Sect.~\ref{subsec:4-2}), and we use the quantum circuits for multi-dimensional cases that are given in Appendix \ref{sec:appA}. 
In both numerical results, we take $N=N_1\times N_2=128\times 128$ and $N_{\text{shot}}=10^5$.
The first function is a trigonometric function given by
$$
f_3(x,y) = \cos\left(2\pi \left((x-0.5)^2+(y-0.5)^2\right)\right) + 1, \quad (x,y)\in [0,1]^2.
$$
The adaptive approximation parameter is $M_1=M_2=32$. The RMSE between the true solution and the RSR reconstruction at the grid points is $3.773$, while the RMSE between the true solution and the FSR reconstruction is $4.216\times 10^{-2}$. 
The second function is a linear combination of two Gaussian functions given by
$$
f_4(x,y) = \exp\left(-25\left((x-0.65)^2+(y-0.65)^2\right)\right) + \exp\left(-16\left((x-0.35)^2+(y-0.35)^2\right)\right), \quad (x,y)\in [0,1]^2.
$$
The adaptive approximation parameter is $M_1=M_2=16$. The RMSE between the true solution and the RSR reconstruction at the grid points is $9.740\times 10^{-1}$, while the RMSE between the true solution and the FSR reconstruction is $1.741\times 10^{-2}$. 
The RMSEs for the FSR method are more than $50$ times smaller than those for the RSR method when the total grid number is as large as $N=2^{14}$. 
We also test the above two functions for a larger grid number $N=2^{18}$ in Fig.~\ref{sec3.5:fig4} and find that the RMSEs for the FSR method do not increase, while the RSR method becomes more than $8$ times worse. This verifies numerically that the proposed FSR method has much better performance in the grid number (i.e., independence of $N$) even for the multi-dimensional cases.

\section{Discussions}
\label{sec4}

In this section, we discuss some variations of the methods in this paper. 
In Sect. \ref{subsec:4-1}, we introduce an overlap-based method based on the QFT that is a similar idea to the one in \cite{Wu.2025pre}. Since this method is not efficient in $N$, we modify it by approximation to obtain an $N$-independent scaling of the cost. However, compared to the FSR method, the pre-factor of the quantum complexity is large (proportional to the number of target points) although the classical complexity is reduced. 
In Sect.~\ref{subsec:4-2}, we propose an adaptive algorithm for the FSR method. Without changing the quantum circuits, we show how to determine the approximation parameter $M$ automatically under a sampling result with a fixed number of repetitions $N_{\text{shot}}$. According to Remark \ref{appC:rem3} in Appendix \ref{sec:appC}, this algorithm has the same theoretical order as that shown in Table \ref{sec3:tab1}, and we regard it as the FSR method for practical applications.  

\subsection{Overlap-based fully quantum Fourier space readout (fQFSR)}
\label{subsec:4-1}

In the above sections, we propose a quantum-classical hybrid method that reads out only the approximations of the Fourier coefficients on quantum computers and reconstructs (the grid-point values of) the underlying function on classical computers. 
In fact, it is possible to reconstruct the function directly on quantum computers by calculating the overlaps $\langle \phi_j |U_{\text{QFT}}^{\dag}\ket{\psi}$, where $U_{\text{QFT}}^{\dag}\ket{\psi}$ gives an amplitude encoding of the Fourier coefficients:
$$
U_{\text{QFT}}^{\dag}\ket{\psi} = \sum_{k=0}^{N-1} c_k\ket{k},
$$
and $\ket{\phi_j}$ is defined by
$$
\ket{\phi_j} := \frac{1}{\sqrt{N}}\sum_{k=0}^{N-1} \mathrm{exp}\left(-\mathrm{i}2\pi kj/N\right)\ket{k}, \quad j=0,\ldots,N-1. 
$$
Noting the discrete Fourier series expansion (see Eq.~\eqref{appC:eq-dfe} in Appendix \ref{sec:appC}) and the relation between the discrete Fourier coefficient and the above (quantum) Fourier coefficient: $c_{k,\text{d}} = \frac{A_N}{\sqrt{N}}c_{k}$, $k=0,\ldots,N-1$, we obtain
$$
\langle \phi_j |U_{\text{QFT}}^{\dag}\ket{\psi} = \frac{1}{\sqrt{N}} \sum_{k=0}^{N-1} c_k \mathrm{exp}\left(\mathrm{i}2\pi kj/N\right) = \frac{f(x_j)}{A_N}, \quad j=0,\ldots,N-1.
$$
Here, the overlap can be calculated by e.g., the Hadamard test or the swap test, but anyway we reach the estimation of the number of repetitions $N_{\text{shot}}=O(N(1/\varepsilon)^2)$ for the error bound $\varepsilon$ due to the scaling $1/A_N \sim 1/\sqrt{N}$ for large $N$. Note that this is the same requirement for the RSR method, which is not satisfactory for large grid numbers. 
Moreover, we mention that by substituting $\ket{\phi_j}$ with 
$$
\ket{\phi_x} := \frac{1}{\sqrt{N}}\sum_{k=0}^{N-1} \mathrm{exp}\left(-\mathrm{i}2\pi kx/L\right)\ket{k}, \quad x\in [0,L],
$$
similarly as the FSR method, we can approximately reconstruct the value of the function even beyond the grid points:
$$
f(x) \approx A_N \langle \phi_x | U_{\text{QFT}}^\dag\ket{\psi}.
$$
The quantum state $\ket{\phi_x}$ can be simply implemented by an Hadamard gate and an $x$-dependent phase gate applied to each qubits. A similar quantum state for the Chebyshev expansion was discussed in \cite{Wu.2025pre,Williams.2023pre}, which is implemented by a non-unitary operation called the Chebyshev feature map for a given $x$. In fact, the readout technique mentioned in \cite{Wu.2025pre} can be regarded as the Chebyshev version of the fQFSR method by calculating the overlaps, and hence, is not efficient for large grid numbers. 

On the other hand, we can follow the idea of the FSR method to improve the overlap-based fully quantum readout method. 
We introduce an approximation parameter $M=O((1/\varepsilon)^{s})<N$ for some $0<s\le 2$. By considering the overlaps between the state for the $M$ dominant Fourier coefficients and the quantum state: 
$$
\ket{\phi_x^M} := \frac{1}{\sqrt{M}}\sum_{k=0}^{M-1} \mathrm{exp}\left(-\mathrm{i}2\pi (k-M/2)x/L\right)\ket{k}, \quad x\in [0,L],
$$ 
we find that the normalized factor is independent of $N$. Then, we obtain the estimation of the number of repetitions: $N_{\text{shot}}=O\left((1/\varepsilon)^{2+s}\right)$, which is much more efficient for large grid numbers. One example of the detailed quantum circuit by the swap test as well as the justification is provided in Appendix \ref{sec:appE}. 
We emphasize that although the fQFSR method can be efficient in $N$ by introducing approximations, the required quantum computations are still expensive since we need to repeat the readout for each target grid point. The proposed FSR method changes the expensive quantum computations into relevant classical computations to achieve further reduction in the total computational cost (as quantum computing is much more expensive than the classical one for the current quantum/classical devices). 

\subsection{Adaptive scheme of determining approximation parameter}
\label{subsec:4-2}

There is a question on the determination of the approximation parameter $M$ in practice. In fact, it is not necessary to know $M$ in advance. By the discussion in Remark \ref{appC:rem3} in Appendix \ref{sec:appC}, the obtained Fourier coefficients after some index related to the number of repetitions $N_{\text{shot}}$ will be automatically zero with a high probability. Ignoring the zero approximate Fourier coefficients gives a natural FSR method with an approximation. 

In other words, we propose the $N_{\text{shot}}$-adaptive algorithm as follows:
\begin{enumerate}
\item Given the number of repetitions $N_{\text{shot}}$.  

\item Apply the quantum circuit in Fig.~\ref{sec2:fig4} and post-select the last (i.e., the most significant) qubit as $\ket{0}$ for $N_{\text{shot}}$ times. Then, we count the the number of obtaining the quantum state $\ket{k}_n$ as $N_{k,\text{sum}}$, which gives the approximate absolute value of the amplitude $d_k = \sqrt{N_{k,\text{sum}}/N_{\text{shot}}} \approx |c_k|$. 

\item Find the index $K_0:=\max\{k=0,1,\ldots,N-1; d_k\not=0\}$, and set $M:=2^{\lceil\log_2 (K_0+1)\rceil}$.  

\item Apply the quantum circuit in Fig.~\ref{sec2:fig5} with the above approximation parameter $M$ to determine the signs of the Fourier coefficients $\{c_k\}_{k=0}^{M-1}$.  

\item Reconstruct the approximate value at any target point $x$ using the formula: 
\begin{align*}
\hat f(x) = \frac{A_N}{\sqrt{N}} \left(\hat c_0 + 2\sum_{k=1}^{M-1} \hat c_k \cos\left(2\pi kx/\tilde L\right)\right), \quad \tilde L = 2L.
\end{align*}
Here, each $\hat c_k$ is the obtained approximation of the Fourier coefficient. 
\end{enumerate}
In the second step, we note that $d_k$ has (nonzero) values only for limited number of indices provided that the repetition number $N_{\text{shot}}$ is relatively small. 
In the third step, we do not consider the indices after $N$ because $|c_k|=|c_{2N-k}|$ for $k=1,\ldots,N$ by the definition of the Fourier coefficients. Moreover, the approximation parameter $M$ is chosen such that $d_k = 0$ for all $k=M,\ldots,N-1$ and there exists $k_0\in [M/2,M-1]$ satisfying $d_{k_0}\not= 0$. 
This implies that we can choose an adaptive $M$ corresponding to a specific sampling result of $N_{\text{shot}}$ times. 
In the fifth step, $\tilde L=2L$ because we have used the (even) extension operator in Figs.~\ref{sec2:fig4} and \ref{sec2:fig5}. 
Moreover, $A_N$ is the normalized factor defined in Sect.~\ref{subsec:2-2}, which is the a priori information. In practice, it can be either derived from the success probability in preparing the input quantum state in the application of the quantum algorithms for differential equations or determined by $A_N = \sqrt{N}f(x^\ast)\left(\hat c_0 + 2\sum_{k=1}^{M-1} \hat c_k \cos\left(2\pi k x^\ast/\tilde L\right)\right)$ if a non-zero value at a single point $x^\ast$ of the underlying function is known. 

Note that for relatively large $N_{\text{shot}}$, it is possible to have some isolated indices $k$ such that $d_k=\sqrt{1/N_{\text{shot}}}$ (i.e., a single shot), which overestimates the necessary $M$. In the case that we intend to minimize $M$, we can introduce a margin parameter $N_{\text{mar}}\in \mathbb{N}$ (e.g., $N_{\text{mar}}=4$) and use the alternative index: 
$$
K_0 := \max\{k=0,1,\ldots,N-1; d_k\not=0 \text{ and }
\left(d_{\max\{k-N_{\text{mar}}, 0\}},\cdots,d_{k-1}, d_{k+1},\cdots,d_{\min\{k+N_{\text{mar}}, N-1\}}\right)\not= \mathbf{0}\},
$$ 
in Step 3. 

\section{Conclusion}
\label{sec5}

We proposed a quantum-classical hybrid readout method called the Fourier space readout (FSR), whose idea is to localize the quantum state by some quantum efficient transforms (e.g., quantum Fourier transform in this paper). The idea can be readily extended to other transforms that can be efficiently implemented on quantum computers. For example, the quantum Chebyshev transform is a candidate, but we need to pay more attention to the gate complexity as well as the success probability of the block encoding since it is a non-unitary operation. 

In particular, for the readout of a real-valued function, we further proposed explicit quantum circuits to efficiently obtain the approximate Fourier coefficients. The quantum complexity, mainly the number of repetitions of the measurements, was confirmed numerically for some examples in Sect.~\ref{sec3}. Moreover, we provided the theoretical supports in Appendix \ref{sec:appC} for the confirmed numerical orders. 
Compared to the real space readout (RSR) method, our proposal has the following strong points (see Table \ref{sec3:tab1}):
\begin{itemize}
\item To achieve a given error bound $\varepsilon$, the FSR method greatly reduces the quantum complexity in the case of a large grid number $N$. Precisely, the number of repetitions of the measurements is independent of the grid number compared to the linear dependence for the RSR method. 

\item In many applications that the underlying function is sufficiently smooth (e.g., absolutely continuous with suitable boundary conditions), the FSR method gives a better performance provided that $N$ is proportional to $(1/\varepsilon)^d$ for $d\ge 1$. Precisely, the dependence on $1/\varepsilon$ is $O\left((1/\varepsilon)^{2+s}\right)$ for $s\le 2/3$, which is polynomially smaller than the dependence $O\left((1/\varepsilon)^{2+d}\right)$ for the RSR method. 

\item Different from the RSR method that the values at all the grid points are derived, the FSR method aims at the features of the underlying function (the Fourier coefficients are important features for continuous functions). Thus, it is more flexible for applications, and the classical computational cost depends on the number of target points, but does not directly depend on the grid number. 
\end{itemize}
Furthermore, we also provide an $N_{\mathrm{shot}}$-adaptive algorithm (Sect.~\ref{subsec:4-2}) to automatically determine the approximation parameter. This indicates that we do not need the a priori information of the underlying functions and is a crucial advantage for real-world applications.  
Our discussion implies that the quantum complexity for the readout of a $(\log_2 N)$-qubit quantum state is not necessarily to have a linear scaling in $N$. By extracting suitable features of the quantum state, the quantum complexity depends only on the desired error bound and the underlying function. Moreover, the classical computational cost for the reconstruction depends only on the number of features and the number of target points, which can also be independent of the grid number $N$ in applications. 

On the other hand, the efficient determinations of the signs of the Fourier coefficients still remain a problem. Although our proposed quantum circuits in Fig.~\ref{sec2:fig5} are effective under numerical experiments in the sense that the mean value of the error is much more smaller than the mean value for the RSR method, the variance of the error is relatively large compared to the variance for the RSR method. This partially comes from the possible inversion of the signs for extremely small Fourier coefficients. This problem does not cancel the promising performance of the FSR method for large grid numbers, but a further efficient way to determine the signs helps to reduce the pre-factor of the repetitions, and hence is expected in practical applications. 


\section*{Acknowledgments}

This work was partially supported by the Center of Innovations for Sustainable Quantum AI (JST Grant number JPMJPF2221).

\appendix

\section{Multi-dimensional FSR method}
\label{sec:appA}

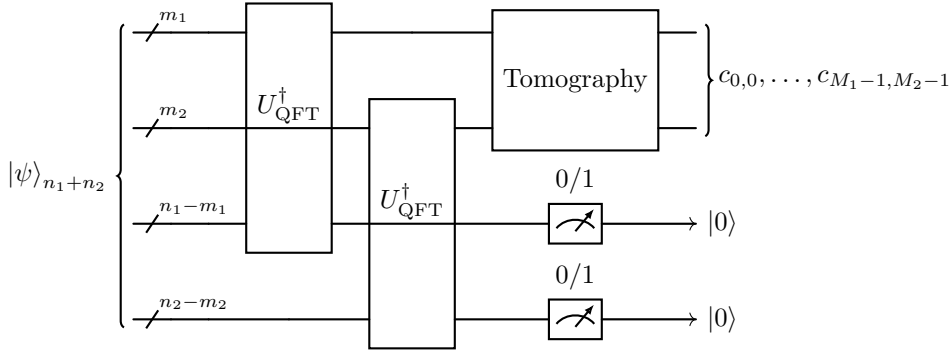
\begin{figure}
\centering
\begin{quantikz}[transparent]
\lstick[4]{$\ket{\psi}_{n_1+n_2}$} & \qwbundle{m_1} &  & \gate[3, label style={yshift=0.3cm}]{U_{\text{QFT}}^\dag} &  & \gate[2]{\text{Tomography}} & \rstick[2]{$c_{0,0},\ldots,c_{M_1-1, M_2-1}$} \\
 & \qwbundle{m_2} &  & \linethrough & \gate[3, label style={yshift=0.3cm}]{U_{\text{QFT}}^\dag} &  &  \\
 & \qwbundle{n_1-m_1} &  &  & \linethrough & \meter{0/1}\arrow[r] &
\rstick{\ket{0}} \\
& \qwbundle{n_2-m_2} &  &  &  & \meter{0/1}\arrow[r] &
\rstick{\ket{0}} 
\end{quantikz}
\caption{A quantum circuit for the 2D Fourier space readout method. Multi-dimensional cases can be discussed in a similar way.}
\label{appA:fig1}
\end{figure}

The one-dimensional FSR method can be readily generalized to the multi-dimensional cases. Let the number of qubits for each dimension be $(n_1,\ldots,n_d)\in \mathbb{N}^d$ where $d$ denotes the dimension. Moreover, choose $(m_1,\ldots,m_d)\in \mathbb{N}^d$ such that $m_\ell \le n_\ell$, and denote $N_\ell = 2^{n_\ell}$, $M_\ell = 2^{m_\ell}$ for $\ell=1,\ldots,d$. Then, the quantum circuit for the multi-dimensional FSR method can be similarly constructed as Fig.~\ref{sec2:fig2}. An example of the two-dimensional case is shown in Fig.~\ref{appA:fig1}. 
Let $(L_1,\ldots,L_d)\in \mathbb{R}^d$ and $f: [0,L_1]\times\cdots\times [0,L_d]\rightarrow \mathbb{R}$ be a multi-variable real-valued function. We introduce the grid points $\mathbf{x}_{\mathbf{j}} = \left(x_{j_1}^{(1)},\ldots,x_{j_d}^{(d)}\right)$ where $x^{(\ell)}_{j_\ell} = j_\ell L_\ell/N_\ell$ for $j_\ell=0,\ldots,N_\ell$ and $\ell=1,\ldots,d$. The normalized factor is defined by
$$
A := \left(\sum_{j_1=0}^{N_1-1}\cdots\sum_{j_d=0}^{N_d-1} \left|f\left(x^{(1)}_{j_1},\ldots, x^{(d)}_{j_d}\right)\right|^2\right)^{1/2},
$$
and the input quantum state is given by
$$
\ket{\psi} = \frac{1}{A} \sum_{j_1=0}^{N_1-1}\cdots\sum_{j_d=0}^{N_d-1} f\left(x^{(1)}_{j_1},\ldots, x^{(d)}_{j_d}\right) \ket{j_d}\otimes \cdots\otimes \ket{j_1}. 
$$
Assume that we obtain the (approximate) Fourier coefficients $c_{j_1,\ldots,j_d}$ for $j_\ell=0,\ldots,M_{\ell}-1$, $\ell=1,\ldots,d$. Then, the function is reconstructed approximately by the following formula:
\begin{equation}
\label{appA:eq-rec}
f\left(x^{(1)},\ldots,x^{(d)}\right) \approx \frac{A}{\prod_{\ell=1}^d \sqrt{N_\ell}} \sum_{j_1=-(M_1-1)}^{M_1-1}\cdots \sum_{j_d=-(M_d-1)}^{M_d-1} c_{j_1,\ldots,j_d} \mathrm{exp}\left(\mathrm{i}\sum_{\ell=1}^d\frac{2\pi j_\ell}{L_\ell}x^{(\ell)}\right).
\end{equation}
Here, the coefficients for negative indices are defined by
$$
c_{j_1,\ldots,j_d}=
\left\{
\begin{aligned}
& c_{|j_1|,\ldots,|j_d|}, && \mbox{if the number of negative indices is even},\\
& \overline{c_{|j_1|,\ldots,|j_d|}}, && \mbox{if the number of negative indices is odd},
\end{aligned}
\right.
$$
where $\overline{c}$ denotes the complex conjugate of a complex number $c$. Under the above definition, one can verify that the right-hand side of Eq.~\eqref{appA:eq-rec} is real-valued for any $\left(x^{(1)},\ldots,x^{(d)}\right) \in [0,L_1]\times \cdots \times [0,L_d]$. 

In the case that the function has the same values at both two boundaries for each dimension (i.e., weak periodic boundary conditions), we can also construct the explicit quantum circuits using simply $Z$-basis measurements which are similar to Figs.~\ref{sec2:fig4} and \ref{sec2:fig5}. The two-dimensional example is provided in Fig.~\ref{appA:fig2}, where $U_{\text{ext},0}$ is the extension operator without the state preparation oracle $U_{\psi}$, see Fig.~\ref{appA:fig3}.  
\begin{figure}
\centering
\resizebox{15cm}{!}{
\begin{quantikz}[transparent]
\lstick{\ket{0}} & \qwbundle{m_1} &  & \gate[5]{U_\psi} & \gate[4]{U_{\text{ext},0}} &  & \gate[4]{U_{\text{QFT}}^\dag} &  & \meter{0/1} & \rstick[2]{$d_{0,0},\ldots,d_{M_1-1,M_2-1}$} \\[-0cm]
\lstick{\ket{0}} & \qwbundle{m_2} &  &  & \linethrough & \gate[5,label style={yshift=-0.3cm}]{U_{\text{ext},0}} & \linethrough & \gate[5,label style={yshift=-0.3cm}]{U_{\text{QFT}}^\dag} & \meter{0/1} &  \\[-0cm]
\lstick{$\ket{0}$} & \qwbundle{n_1-m_1} &  &  &  & \linethrough &  & \linethrough
 & \meter{0/1}\arrow[r] &
\rstick{\ket{0}} \\
\lstick{\ket{0}} & \qw &  & \linethrough &  & \linethrough &  & \linethrough & \meter{0/1}\arrow[r] &
\rstick{\ket{0}} \\
\lstick{$\ket{0}$} & \qwbundle{n_2-m_2} &  &  &  &  &  &  & \meter{0/1}\arrow[r] &
\rstick{\ket{0}} \\
\lstick{\ket{0}} & \qw &  &  &  &  &  &  & \meter{0/1}\arrow[r] &
\rstick{\ket{0}} 
\end{quantikz}
\qquad
\begin{quantikz}[transparent]
\lstick{\ket{0}} & \qwbundle{m_1} & \gate[5]{U_\psi} & \gate[4]{U_{\text{ext},0}} &  & \gate[4]{U_{\text{QFT}}^\dag} &  & \gate[2]{H^{\otimes (m_1+m_2)}} &  & \meter{0/1} & \rstick[2]{$e_{0,0},\ldots,e_{M_1-1,M_2-1}$} \\[-0cm]
\lstick{\ket{0}} & \qwbundle{m_2} &  & \linethrough & \gate[5,label style={yshift=-0.3cm}]{U_{\text{ext},0}} & \linethrough & \gate[5]{U_{\text{QFT}}^\dag} &  &  & \meter{0/1} &  \\[-0cm]
\lstick{\ket{0}} & \qwbundle{n_1-m_1} &  &  & \linethrough &  &  &  &  & \meter{0/1}\arrow[r] &
\rstick{\ket{0}} \\[-0cm]
\lstick{\ket{0}} &  & \linethrough &  & \linethrough &  &  &  &  & \meter{0/1}\arrow[r] & \rstick{\ket{0}} \\[-0cm]
\lstick{\ket{0}} & \qwbundle{n_2-m_2} &  &  &  &  &  &  &  & \meter{0/1}\arrow[r] &
\rstick{\ket{0}} \\
\lstick{\ket{0}} & \qw &  &  &  &  &  &  &  & \meter{0/1}\arrow[r] &
\rstick{\ket{0}} \\
\lstick{\ket{0}} & \gate{H} & \ctrl{-2} & \ctrl{-3} & \ctrl{-1} & \ctrl{-3} & \ctrl{-1} & \octrl{-5} & \gate{H} & \meter{0/1}\arrow[r] &
\rstick{\ket{0}} 
\end{quantikz}
}
\caption{Explicit quantum circuits using $Z$-basis measurements for the FSR method in the 2D case. Multi-dimensional cases can be discussed in a similar way.}
\label{appA:fig2}
\end{figure}
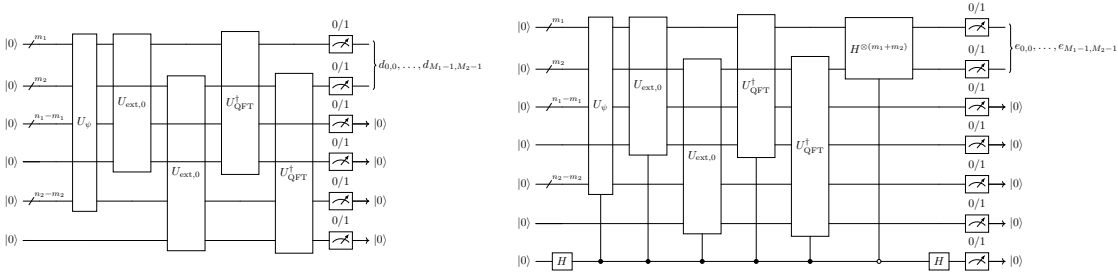
\begin{figure}
\centering
\begin{quantikz}
\lstick[4]{$\ket{0}_n$} &  \gate[5]{U_{\text{ext},0}} & \\
&  & \\
\setwiretype{n} & \vdots &  \\
&  & \\
\lstick{\ket{0}} & \qw &  
\end{quantikz}
= 
\begin{quantikz}
&  & \targ{} & \gate[4]{U_{+1}} & \\[-0cm]
&  & \targ{} &  & \\[-0cm]
\setwiretype{n} & \vdots & \vdots &  & \\
&  & \targ{} &  & \\[-0cm]
& \gate{H} & \ctrl{-4} & \ctrl{-1} & 
\end{quantikz}
\caption{A quantum circuit for an extension operator $U_{\text{ext},0}$. Here, $U_{+1}$ denotes the quantum incrementer gate.}
\label{appA:fig3}
\end{figure}
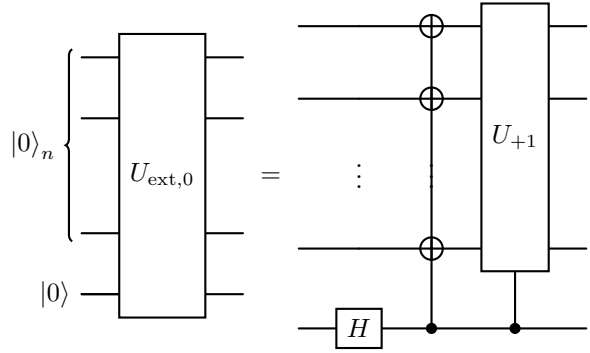

\section{Real-valued Fourier coefficients by even extension}
\label{sec:appB}

The quantum state with all real-valued amplitudes can be obtained with simply $Z$-basis measurements for all the qubits. To obtain real-valued Fourier coefficients, we introduced the even extension operator $U_{\text{ext},\psi}$ in Sect.~\ref{subsec:2-3}, which used the oracle $U_\psi$, an Hadamard gate, a couple of CNOT gates, and a controlled incrementer gate (see Fig.~\ref{sec2:fig3}). 

With the Hadamard gate and the input oracle, the initialized quantum state becomes
$$
\ket{\psi_1} = \frac{1}{\sqrt{2}}(\ket{0}+\ket{1}) \otimes \frac{1}{A_N}\left(\sum_{j=0}^{N-1} f(x_j)\ket{j}_n\right).  
$$
Then, the series of CNOTs invert the first $n$ qubits if the last qubit is in $\ket{1}$, which yields
$$
\ket{\psi_2} = \frac{1}{\sqrt{2}A_N} \sum_{j=0}^{N-1} f(x_j)\ket{0} \otimes \ket{j}_n + \frac{1}{\sqrt{2}A_N} \sum_{j=0}^{N-1} f(x_j) \ket{1} \otimes \ket{N-1-j}_n.
$$
Finally, the controlled incrementer gate changes the quantum state into
\begin{align*}
\ket{\psi_3} &= \frac{1}{\sqrt{2}A_N} \sum_{j=0}^{N-1} f(x_j)\ket{0} \otimes \ket{j}_n + \frac{1}{\sqrt{2}A_N} \sum_{j=0}^{N-1} f(x_j) \ket{1} \otimes \ket{N-j}_n \\
&= \frac{1}{\sqrt{2}A_N} \sum_{j=0}^{N-1} f(x_j)\ket{0} \otimes \ket{j}_n + \frac{1}{\sqrt{2}A_N} \sum_{j=1}^{N} f(x_{N-j}) \ket{1} \otimes \ket{j}_n \\
&=  \frac{1}{\sqrt{2}A_N} \sum_{j=0}^{N-1} f(x_j)\ket{0} \otimes \ket{j}_n + \frac{1}{\sqrt{2}A_N} \sum_{j=0}^{N-1} f(x_{N-j}) \ket{1} \otimes \ket{j}_n.
\end{align*}
In the last equality, we used $f(x_0) = f(x_N)$ which comes from the assumption $f(0)=f(L)$. Thus, we verified Eq.~\eqref{sec2:eq-ext}, and construct a quantum state corresponding to an even extension. 

Next, we confirm that the quantum state after applying the inverse QFT has all real-valued amplitudes. After the extension operator, we obtain a center-symmetric function. So in general, we consider a function $\tilde f: [0,\tilde L]\rightarrow \mathbb{R}$ satisfying $\tilde f(\tilde x) = \tilde f(\tilde L- \tilde x)$ for $\tilde x\in [0,\tilde L]$. With an integer $\tilde n\in \mathbb{N}$ and $\tilde N = 2^{\tilde n}$, we introduce the grid points $\tilde x_j = j\tilde L/\tilde N$, $j=0,\ldots,\tilde N$. 
Moreover, the normalized factor is defined by 
$$
\tilde A_{\tilde N} := \left(\sum_{j=0}^{\tilde N-1} |\tilde f(\tilde x_j)|^2\right)^{1/2}.
$$
Given a quantum state: 
$$
\ket{\tilde \psi} = \frac{1}{\tilde A_{\tilde N}}\sum_{j=0}^{\tilde N-1} \tilde f(\tilde x_j) \ket{j}_{\tilde n}.
$$
By the definition, we have 
\begin{align*}
U_{\text{QFT}}^\dag \ket{\tilde \psi} &= \frac{1}{\sqrt{\tilde N}\tilde{A}_{\tilde N}} \sum_{k=0}^{\tilde N-1} \tilde f(\tilde x_k) \sum_{j=0}^{\tilde N-1} \mathrm{exp}\left(-\mathrm{i}\frac{2\pi}{\tilde N}kj\right) \ket{j}_{\tilde n} \\
&= \sum_{j=0}^{\tilde N-1} \left(\frac{1}{\sqrt{\tilde N}\tilde{A}_{\tilde N}} \sum_{k=0}^{\tilde N-1} \tilde f(\tilde x_k) \mathrm{exp}\left(-\mathrm{i}\frac{2\pi}{\tilde N}kj\right) \right) \ket{j}_{\tilde n} =: \sum_{j=0}^{\tilde N-1} \tilde c_j \ket{j}_{\tilde n}. 
\end{align*}
Thus, using the symmetric condition: $\tilde f(\tilde x_j) = \tilde f(\tilde x_{\tilde N-j})$ for $j=0,\ldots,\tilde N$, we calculate
\begin{align*}
\sqrt{\tilde N}\tilde{A}_{\tilde N} \tilde c_j &= \sum_{k=0}^{\tilde N-1} \tilde f(\tilde x_k) \mathrm{exp}\left(-\mathrm{i}\frac{2\pi}{\tilde N}kj\right) \\
&= \sum_{k=0}^{\tilde N-1} \tilde f(\tilde x_k) \cos\left(\frac{2\pi}{\tilde N}kj\right) -\mathrm{i}\sum_{k=0}^{\tilde N-1} \tilde f(\tilde x_k) \sin\left(\frac{2\pi}{\tilde N}kj\right) \\
&= \sum_{k=0}^{\tilde N-1} \tilde f(\tilde x_k) \cos\left(\frac{2\pi}{\tilde N}kj\right) -\frac{\mathrm{i}}{2}\sum_{k=0}^{\tilde N-1} \tilde f(\tilde x_k) \sin\left(\frac{2\pi}{\tilde N}kj\right) -\frac{\mathrm{i}}{2}\sum_{k=0}^{\tilde N-1} \tilde f(\tilde x_{k}) \sin\left(\frac{2\pi}{\tilde N}k j\right)\\
&= \sum_{k=0}^{\tilde N-1} \tilde f(\tilde x_k) \cos\left(\frac{2\pi}{\tilde N}kj\right) -\frac{\mathrm{i}}{2}\sum_{k=0}^{\tilde N-1} \tilde f(\tilde x_k) \sin\left(\frac{2\pi}{\tilde N}kj\right) -\frac{\mathrm{i}}{2}\sum_{k^\prime=1}^{\tilde N} \tilde f(\tilde x_{\tilde N-k^\prime}) \sin\left(\frac{2\pi}{\tilde N}(\tilde N-k^\prime) j\right) \\
&= \sum_{k=0}^{\tilde N-1} \tilde f(\tilde x_k) \cos\left(\frac{2\pi}{\tilde N}kj\right) -\frac{\mathrm{i}}{2}\sum_{k=1}^{\tilde N-1} \left(\tilde f(\tilde x_k) - \tilde f(\tilde x_{\tilde N-k})\right) \sin\left(\frac{2\pi}{\tilde N}kj\right) \\
&= \sum_{k=0}^{\tilde N-1} \tilde f(\tilde x_k) \cos\left(\frac{2\pi}{\tilde N}kj\right) \in \mathbb{R}.
\end{align*}
In the second last line, we used $\sin(0)=0$ and $\sin(2\pi j-\tilde x)=-\sin(\tilde x)$. We mention that for multi-dimensional cases, the quantum state still has all real-valued amplitudes as long as we assume similar symmetric structures for each dimension. 

In this paper, we call the above Fourier coefficients $\tilde c_j$ the quantum Fourier coefficients and denote it by $\tilde c_{j,\text{q}}$ if we need to clarify it. Moreover, the discrete Fourier coefficients and the continuous Fourier coefficients of $\tilde f$ are defined by
\begin{equation}
\label{appB:eq-dfc}
\tilde c_{j, \text{d}} := \frac{1}{\tilde N} \sum_{k=0}^{\tilde N-1} \tilde f(\tilde x_k) \mathrm{exp}\left(-\mathrm{i}\frac{2\pi}{\tilde N}kj\right), \quad j=0,\ldots,\tilde N-1,
\end{equation}
\begin{equation}
\tilde c_{j, \text{c}} := \frac{1}{\tilde L} \int_{0}^{\tilde L} \tilde f(\tilde x) \mathrm{exp}\left(-\mathrm{i}\frac{2\pi j}{\tilde L}\tilde x\right) \mathrm{d}\tilde x, \quad j\in \mathbb{Z},
\end{equation}
so that the discrete and continuous Fourier series expansions are given by
\begin{equation*}
\tilde f(\tilde x_k) = \sum_{j=0}^{\tilde N-1} \tilde c_{j, \text{d}} \mathrm{exp}\left(\mathrm{i}\frac{2\pi}{\tilde N}kj\right), \quad k=0,\ldots,\tilde N-1,
\end{equation*}
\begin{equation*}
\tilde f(\tilde x) = \sum_{j=-\infty}^\infty \tilde c_{j, \text{c}} \mathrm{exp}\left(\mathrm{i}\frac{2\pi j}{\tilde L}\tilde x\right), \quad \tilde x\in [0,\tilde L]. 
\end{equation*}
Therefore, the relation between quantum Fourier coefficients and the discrete/continuous Fourier coefficients is given as follows:
\begin{align}
\nonumber
\tilde c_{j,\text{q}} &= \frac{1}{\sqrt{\tilde N}\tilde{A}_{\tilde N}} \sum_{k=0}^{\tilde N-1} \tilde f(\tilde x_k) \mathrm{exp}\left(-\mathrm{i}\frac{2\pi}{\tilde N}kj\right) = \frac{\sqrt{\tilde N}}{\tilde A_{\tilde N}} \tilde c_{j,\text{d}} \\
\label{appB:eq-rqc}
&= \frac{\sqrt{\tilde N}}{\tilde A_{\tilde N}} \frac{1}{\tilde L} \sum_{k=0}^{\tilde N-1} \tilde f(\tilde x_k) \mathrm{exp}\left(-\mathrm{i}\frac{2\pi j}{\tilde L}\tilde x_k \right) \frac{\tilde L}{\tilde N} \xrightarrow{\tilde N\to \infty} \frac{\sqrt{\tilde L}}{\|\tilde f\|_{L^2}} \tilde c_{j,\text{c}}, \quad j=0,\ldots,\tilde N-1.
\end{align}
Recalling that $\tilde A_{\tilde N} = O\left(\sqrt{\tilde N}\right)$, all of the above three types of Fourier coefficients have uniform dependence on $\tilde N$. 
We can approximately reconstruct the function by 
\begin{align}
\nonumber
\tilde f(\tilde x) &\approx \sum_{j=-(M-1)}^{M-1} \tilde c_{j, \text{c}} \mathrm{exp}\left(\mathrm{i}\frac{2\pi j}{\tilde L}\tilde x\right) 
\approx \sum_{j=-(M-1)}^{M-1} \tilde c_{j, \text{d}} \mathrm{exp}\left(\mathrm{i}\frac{2\pi j}{\tilde L}\tilde x\right) \\
\nonumber
&= \frac{\tilde A_{\tilde N}}{\sqrt{\tilde N}}\sum_{j=-(M-1)}^{M-1} \tilde c_{j, \text{q}} \mathrm{exp}\left(\mathrm{i}\frac{2\pi j}{\tilde L}\tilde x\right)\\
\label{appB:eq-rec}
&= \frac{\tilde A_{\tilde N}}{\sqrt{\tilde N}} \left(\tilde c_{0,\text{q}} + 2\sum_{k=1}^{M-1} \mathcal{R}\left(\tilde c_{k, \text{q}} \mathrm{exp}\left(\mathrm{i}\frac{2\pi k}{\tilde L}\tilde x\right)\right) \right),
\end{align}
for an integer $M\le \tilde N$, which is chosen regarding the desired error bound.  

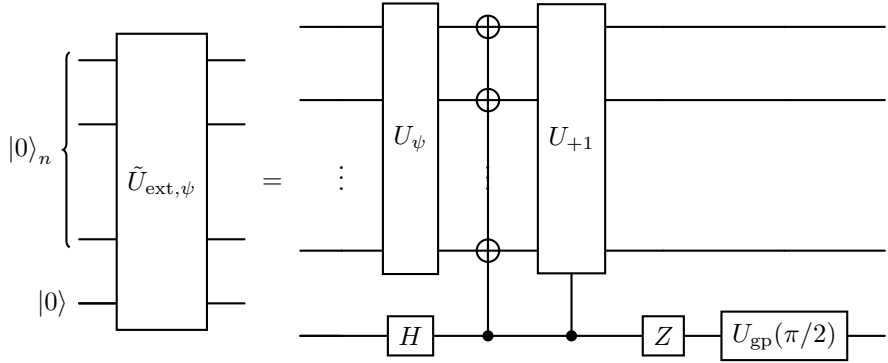
\begin{figure}
\centering
\begin{quantikz}
\lstick[4]{$\ket{0}_n$} &  \gate[5]{\tilde{U}_{\text{ext},\psi}} & \\
&  & \\
\setwiretype{n} & \vdots &  \\
&  & \\
\lstick{\ket{0}} & \qw &  
\end{quantikz}
= 
\begin{quantikz}
&  & \gate[4]{U_\psi} & \targ{} & \gate[4]{U_{+1}} &  &  &  \\[-0cm]
&  &  & \targ{} &  &  &  &  \\[-0cm]
\setwiretype{n} & \vdots &  & \vdots &  &  &  &  \\
&  &  & \targ{} &  &  &  &  \\[-0cm]
& \qw & \gate{H} & \ctrl{-4} & \ctrl{-1} & \gate{Z} & \gate{U_{\text{gp}}(\pi/2)} &  
\end{quantikz}
\caption{A quantum circuit for the odd extension operator $\tilde{U}_{\text{ext},\psi}$. Here, $U_{+1}$ and $U_{\text{gp}}(\pi/2)$ denote the quantum incrementer gate and a global phase gate with the phase rotation of $\pi/2$ radians, respectively. }
\label{appC:fig1}
\end{figure}

As mentioned in Sect.~\ref{subsec:2-3}, we can also employ one ancillary qubit and introduce the following odd extension operator:
\begin{equation}
\label{appC:eq-ext-odd}
\tilde{U}_{\text{ext},\psi} \left(\ket{0} \otimes \ket{0}_n\right) = \frac{\mathrm{i}}{\sqrt{2}A_N}\left(\sum_{j=0}^{N-1} f(x_j)\ket{j}_{n+1} - \sum_{j=0}^{N-1} f(x_{N-j})\ket{N+j}_{n+1}\right).
\end{equation}
With the operation of $U_{\text{QFT}}^\dag$ after the above extension operator, the Fourier coefficients are again real-valued by similar calculations as above. Such an odd extension is especially efficient in the case that the boundary values are zero, and the derivatives at the boundary are opposite. 
Assuming $f(x_0)=f(x_N)=0$, the above extension operator can be realized by the quantum circuit shown in Fig.~\ref{appC:fig1}, which applies an additional single-qubit gate after the quantum circuit for the even extension (see Fig.~\ref{sec2:fig3}). Here, $Z$ is the Pauli-Z gate, and $U_{\text{gp}}(\pi/2)$ denotes a global phase gate with the phase rotation of $\pi/2$ radians. For example, the global phase gate can be implemented by $P(\pi/2)XP(\pi/2)X$ where $X$ is the Pauli-X gate, and $P(\theta)$ is a phase gate with rotation angle $\theta$. 

\section{Theoretical details}
\label{sec:appC}

\subsection{Evaluation of approximation parameter}
\label{subsec:C-1}

Let $f$: $[0,L]\to \mathbb{R}$ be a real-valued function and $f$ has the following (continuous) Fourier series expansion:
$$
f(x) = \sum_{k=-\infty}^\infty c_{k,\text{c}} \mathrm{exp}\left(\mathrm{i}(2\pi/L) kx\right). 
$$
Let $N=2^n$ be the number of grid points. Taking $x=x_j$, where $x_j=jL/N$, $j=0,\ldots,N$ denote the grid points, we have
$$
f(x_j) = \sum_{k=-\infty}^\infty c_{k,\text{c}} \mathrm{exp}\left(\mathrm{i}(2\pi/L) k x_j\right)
= \sum_{k=-\infty}^\infty c_{k,\text{c}} \mathrm{exp}\left(\mathrm{i}2\pi kj/N\right), \quad j=0,\ldots,N-1.
$$
If we truncate the infinite series by $N/2$, we rewrite the above equation by 
\begin{equation}
\label{appC:eq-trcfe}
f(x_j) = \sum_{k=-N/2}^{N/2-1} c_{k,\text{c}} \mathrm{exp}\left(\mathrm{i}2\pi kj/N\right) + R_j, \quad j=0,\ldots,N-1, 
\end{equation}
where $R_j = R_j(N):= \left(\sum_{k=-\infty}^{-N/2-1} + \sum_{k=N/2}^{\infty}\right) c_{k,\text{c}} \mathrm{exp}\left(\mathrm{i}2\pi kj/N\right)$. For the grid method, we have the values of $f$ at $N$ grid points, then the discrete Fourier series expansion yields
\begin{equation}
\label{appC:eq-dfe}
f(x_j) = \sum_{k=0}^{N-1} c_{k,\text{d}} \mathrm{exp}\left(\mathrm{i}2\pi kj/N\right), \quad j=0,\ldots,N-1. 
\end{equation}
Here, $c_{k,\text{d}}$, $k=0,\ldots,N-1$ are the discrete Fourier coefficients. As we discussed in Appendix \ref{sec:appB}, the quantum Fourier coefficients defined by 
\begin{equation}
\label{appC:eq-qfc}
c_{k,\text{q}} := \frac{1}{\sqrt{N}A_N} \sum_{j=0}^{N-1} f(x_j) \mathrm{exp}\left(-\mathrm{i}2\pi kj/N\right), \quad k=0,\ldots,N-1,
\end{equation}
can be approximately derived using the quantum circuits in Fig.~\ref{sec2:fig2} or Figs.~\ref{sec2:fig4} and \ref{sec2:fig5}. Here, $A_N$ is the normalized factor defined by
$$
A_N := \left(\sum_{j=0}^{N-1}|f(x_j)|^2\right)^{1/2}.
$$
Moreover, noting the definition of the discrete Fourier coefficients (see Eq.~\eqref{appB:eq-dfc}), we have the relation $c_{k,\text{d}} = \frac{A_N}{\sqrt{N}}c_{k,\text{q}}$ for $k=0,\ldots,N-1$. 
According to the definitions of discrete/quantum Fourier coefficients, we have $c_{k,\text{d}}=\overline{c_{N-k,\text{d}}}$ and $c_{k,\text{q}}=\overline{c_{N-k,\text{q}}}$ for $k=1,\ldots,N-1$. Here, $\overline{\cdot}$ denotes the complex conjugate of a complex number. Thus, Eq.~\eqref{appC:eq-dfe} yields
\begin{align*}
f(x_j) &= \sum_{k=0}^{N/2-1} c_{k,\text{d}} \mathrm{exp}\left(\mathrm{i}2\pi kj/N\right) + \sum_{k=N/2}^{N-1} c_{k,\text{d}} \mathrm{exp}\left(\mathrm{i}2\pi kj/N\right) \\
&= \sum_{k=0}^{N/2-1} c_{k,\text{d}} \mathrm{exp}\left(\mathrm{i}2\pi kj/N\right) + \sum_{k=N/2}^{N-1} \overline{c_{N-k,\text{d}}} \mathrm{exp}\left(\mathrm{i}2\pi (k-N)j/N\right) \\
&= \sum_{k=0}^{N/2-1} c_{k,\text{d}} \mathrm{exp}\left(\mathrm{i}2\pi kj/N\right) + \sum_{k=-N/2}^{-1} \overline{c_{-k,\text{d}}} \mathrm{exp}\left(\mathrm{i}2\pi kj/N\right).
\end{align*}
By the definition of $c_{k,\text{d}}$ in Eq.~\eqref{appB:eq-dfc}, we can formally extend the Fourier coefficients for negative indices $k=-N/2,\ldots,-1$ with the same formula, and we have $c_{k,\text{d}} = \overline{c_{-k,\text{d}}}$ for $k=0,\ldots,N-1$ since $f$ is real-valued. Together with the above equality, we obtain 
\begin{align}
\label{appC:eq-dfe2}
f(x_j) = \sum_{k=-N/2}^{N/2-1} c_{k,\text{d}} \mathrm{exp}\left(\mathrm{i}2\pi kj/N\right), \quad j=0,\ldots,N-1.
\end{align}
Thus, for a sufficiently large $M\le N/2$, we reconstruct the function at each grid point $x_j$ by 
\begin{align}
\label{appC:eq-arecf}
f_{j,M} := \sum_{k=-(M-1)}^{M-1} c_{k,\text{d}} \mathrm{exp}\left(\mathrm{i}2\pi kj/N\right)
= \frac{A_N}{\sqrt{N}} \left(\sum_{k=-(M-1)}^{M-1} c_{k,\text{q}} \mathrm{exp}\left(\mathrm{i}2\pi kj/N\right)\right).
\end{align}
Here, the quantum Fourier coefficients are also naturally extended for negative indices. One can readily check that the right-hand side above is equivalent to the right-hand side of Eq.~\eqref{appB:eq-rec} since $c_{k,\text{q}}=\overline{c_{-k,\text{q}}}$ for $k=-(M-1),\ldots,-1$. Next, we estimate the squared $l^2$ norm as follows:
\begin{align*}
\sum_{j=0}^{N-1}|f(x_j)-f_{j,M}|^2 
&= \sum_{j=0}^{N-1}\left(\sum_{k\in \mathcal{K}}c_{k,\text{d}}\mathrm{exp}\left(\mathrm{i}2\pi kj/N\right)\right) \overline{\left(\sum_{k^\prime\in \mathcal{K}}c_{k^\prime,\text{d}}\mathrm{exp}\left(\mathrm{i}2\pi k^\prime j/N\right)\right)} \\
&= \sum_{k,k^\prime\in \mathcal{K}} c_{k,\text{d}}\overline{c_{k^\prime,\text{d}}} \sum_{j=0}^{N-1} \mathrm{exp}\left(\mathrm{i}2\pi (k-k^\prime) j/N\right) = N \sum_{k\in \mathcal{K}} |c_{k,\text{d}}|^2.
\end{align*}
Here, $\mathcal{K} := \{-N/2,-N/2+1\ldots,-M, M,M+1,\ldots,N/2-1\}$, and we used the identity: 
$$
\sum_{j=0}^{N-1} \mathrm{exp}\left(\mathrm{i}2\pi (k-k^\prime) j/N\right) = N \delta_{k,k^\prime},
$$
where $\delta_{k,k^\prime}$ is the Kronecker delta. Therefore, the root mean square error (RMSE) and the $l^2$ error for the normalized solution (L2NS) can be estimated as follows:
\begin{align*}
& \text{Err}_{\text{RMSE}} = \left(\frac{1}{N}\sum_{j=0}^{N-1}|f(x_j)-f_{j,M}|^2\right)^{1/2} = \left(\sum_{k\in \mathcal{K}} |c_{k,\text{d}}|^2\right)^{1/2} \le \left(2\sum_{k=M}^{N/2} |c_{k,\text{d}}|^2\right)^{1/2},\\
& \text{Err}_{\text{L2NS}} = \left(\sum_{j=0}^{N-1}\left|\frac{f(x_j)-f_{j,M}}{A_N}\right|^2\right)^{1/2} = \frac{\sqrt{N}}{A_N} \left(\sum_{k\in \mathcal{K}} |c_{k,\text{d}}|^2\right)^{1/2} \le \frac{\sqrt{N}}{A_N}\left(2\sum_{k=M}^{N/2} |c_{k,\text{d}}|^2\right)^{1/2}.
\end{align*}
Finally, we need to estimate $c_{k,\text{d}}$ for $k=M,\ldots,N/2-1$. 
If $f$ has the continuous Fourier expansion exactly, then according to Eq.~\eqref{appC:eq-trcfe} and Eq.~\eqref{appC:eq-dfe2}, we have
\begin{align*}
c_{k,\text{d}}-c_{k,\text{c}} &= \frac{1}{N} \sum_{j=0}^{N-1} R_j \mathrm{exp}\left(-\mathrm{i}2\pi kj/N\right) \\
&= \frac{1}{N}\left(\sum_{k^\prime=\infty}^{-N/2-1} + \sum_{k^\prime=N/2}^{\infty}\right) c_{k^\prime,\text{c}} \sum_{j=0}^{N-1} \mathrm{exp}\left(-\mathrm{i}2\pi (k-k^\prime)j/N\right) = 0,
\end{align*}
for $k=-N/2,\ldots, N/2-1$. This implies that it is sufficient to estimate the continuous Fourier coefficients, which is, to some extent, well-known (e.g., Lemma 17 in Childs et al. \cite{Childs.2022}). Here, we make a detailed estimation according to our application as the following lemma.
\begin{lemma}
\label{appC:lem1}
Let $f$ be a piecewise $W^{1,1}$ function on $[0,L]$, that is, there exists $S\in \mathbb{N}$ and a partition $\Pi$: $0=y_0<y_1<\cdots<y_{S}=L$ such that $f(x)=f_s(x)$, $x\in I_s := (y_{s-1},y_{s})$ for $s=1,\ldots,S$ and $f_s\in W^{1,1}(I_s)$. Then, the Fourier coefficients satisfy
$$
|c_{k,\text{c}}| \le \frac{V_{\Pi}(f)+\left\|f^{(1)}\right\|_{\Pi}}{2\pi k}=O(k^{-1}), \quad k\in \mathbb{Z},
$$
where $V_{\Pi}(f)$ and $\left\|f^{(1)}\right\|_{\Pi}$ are given in the proof. 
Moreover, let $f^{(q)}$ be the $q$-th weak derivative of $f$ and assume $f\in W^{p,1}(0,L)$ and $f^{(q)}(0) = f^{(q)}(L)$, $q=0,\ldots,p_0-2$ for some integers $p\ge p_0\ge 1$. In the case of $p_0=1$, this indicates no boundary assumptions. Then, the Fourier coefficients satisfy
$$
|c_{k,\text{c}}| \le \frac{1}{L} \sum_{q=p_0-1}^{p-1} \left(\frac{L}{2\pi k}\right)^{q+1}\left|f^{(q)}(L)-f^{(q)}(0)\right| + \frac{1}{L} \left(\frac{L}{2\pi k}\right)^{p} \left\|f^{(p)}\right\|_{L^1(0,L)} = O(k^{-p_0}), \quad k\in \mathbb{Z}.
$$
\end{lemma}
\noindent Here and henceforth, $W^{k,p}$ denotes the Sobolev space \cite{Adams.2003}, which indicates the set of functions whose (up to) $k$-th weak derivatives belong to the Lebesgue space $L^p$. In particular, $W^{1,1}$ is known as the space for the absolutely continuous functions. 
\begin{proof}
We calculate by the definition and integration by parts. For simplicity, we use $c_{k}$ instead of $c_{k,\text{c}}$ in the proof. 
For the general case of a piecewise $W^{1,1}$ function, we have
\begin{align*}
c_k &= \frac{1}{L}\int_{0}^L f(x) \mathrm{exp}\left(-\mathrm{i}2\pi kx/L \right) \mathrm{d}x = \frac{1}{L} \sum_{s=1}^S \int_{I_s} f_s(x) \mathrm{exp}\left(-\mathrm{i}2\pi kx/L \right) \mathrm{d}x \\
&= \frac{1}{-\mathrm{i}2\pi k} \sum_{s=1}^S \left(f_s(y_{s}) \mathrm{exp}\left(-\mathrm{i}2\pi ky_{s}/L \right) - f_s(y_{s-1}) \mathrm{exp}\left(-\mathrm{i}2\pi ky_{s-1}/L \right) - \int_{I_s} f_s^{(1)}(x) \mathrm{exp}\left(-\mathrm{i}2\pi kx/L \right) \mathrm{d}x\right)\\
&= \frac{\mathrm{i}}{2\pi k} \left(f_S(L)-f_1(0) + \sum_{s=1}^{S-1} (f_s(y_s)-f_{s+1}(y_s))\mathrm{exp}\left(-\mathrm{i}2\pi ky_s/L\right) - \sum_{s=1}^S\int_{I_s} f_s^{(1)}(x) \mathrm{exp}\left(-\mathrm{i}2\pi kx/L \right) \mathrm{d}x\right). 
\end{align*}
By introducing 
$$
V_{\Pi}(f) := |f_S(L)-f_1(0)| + \sum_{s=1}^{S-1} |f_s(y_s)-f_{s+1}(y_{s})|, \quad \|f^{(1)}\|_{\Pi} := \sum_{s=1}^{S} \left\|f_s^{(1)}\right\|_{L^1(I_s)},
$$
we estimate
$$
|c_k| \le \frac{V_{\Pi}(f)+\left\|f^{(1)}\right\|_{\Pi}}{2\pi k},
$$
for $k\in \mathbb{Z}$. Moreover, if $f\in W^{p,1}(0,L)$, then we can further use integration by parts $p-1$ times to obtain
\begin{align*}
c_k &= \frac{1}{-\mathrm{i} 2\pi k} \left(f(L)-f(0) - \int_0^L f^{(1)}(x) \mathrm{exp}\left(-\mathrm{i}2\pi kx/L \right) \mathrm{d}x\right)\\
&= \frac{1}{-\mathrm{i} 2\pi k}(f(L)-f(0)) -\frac{1}{L} \left(\frac{L}{-\mathrm{i}2\pi k}\right)^2\left(f^{(1)}(L)-f^{(1)}(0) - \int_0^L f^{(2)}(x) \mathrm{exp}\left(-\mathrm{i}2\pi kx/L \right) \mathrm{d}x\right) \\
&= \cdots = -\frac{1}{L}\sum_{q=0}^{p-1} \left(\frac{L}{\mathrm{i}2\pi k}\right)^{q+1}\left(f^{(q)}(L)-f^{(q)}(0)\right) + \frac{1}{L} \left(\frac{L}{\mathrm{i}2\pi k}\right)^{p} \int_0^L f^{(p)}(x) \mathrm{exp}\left(-\mathrm{i}2\pi kx/L \right) \mathrm{d}x\\
&= -\frac{1}{L}\sum_{q=p_0-1}^{p-1} \left(\frac{L}{\mathrm{i}2\pi k}\right)^{q+1}\left(f^{(q)}(L)-f^{(q)}(0)\right) + \frac{1}{L} \left(\frac{L}{\mathrm{i}2\pi k}\right)^{p} \int_0^L f^{(p)}(x) \mathrm{exp}\left(-\mathrm{i}2\pi kx/L \right) \mathrm{d}x
\end{align*}
In the last line above, we used the assumption that $f^{(q)}(0) = f^{(q)}(L)$, $q=0,\ldots,p_0-2$. Therefore, we have 
\begin{align*}
|c_k| \le \frac{1}{L} \sum_{q=p_0-1}^{p-1} \left(\frac{L}{2\pi k}\right)^{q+1}\left|f^{(q)}(L)-f^{(q)}(0)\right| + \frac{1}{L} \left(\frac{L}{2\pi k}\right)^{p} \left\|f^{(p)}\right\|_{L^1(0,L)}.
\end{align*}
In particular, if $p_0=p$, then we have
\begin{align*}
|c_k| \le \frac{1}{L} \left(\frac{L}{2\pi k}\right)^{p} \left(\left|f^{(p-1)}(L)-f^{(p-1)}(0)\right| +  \left\|f^{(p)}\right\|_{L^1(0,L)}\right). 
\end{align*}
\end{proof}
\begin{remark}[Sharpness of the order]
\label{appC:rem1}
Lemma \ref{appC:lem1} shows that for a piecewise differentiable function, the continuous Fourier coefficients have the order $O(k^{-1})$ regarding $k$, and this order is sharp if there are discontinuous points (e.g., piecewise constants). 
Besides, if we assume that $f\in W^{p,1}(0,L)$ satisfying the boundary conditions $f^{(q)}(0)=f^{(q)}(L)$, $q=0,\ldots,p-2$ for $p\ge 1$, then the Fourier coefficients have the order $O(k^{-p})$. Similarly, this order is sharp provided that $f^{(p-1)}(0)\not=f^{(p-1)}(L)$.  
\end{remark}
Combining the above discussions, we have 
\begin{align*}
\left(2\sum_{k=M}^{N/2} |c_{k,\text{d}}|^2\right)^{1/2} &= \left(2\sum_{k=M}^{N/2} |c_{k,\text{c}}|^2\right)^{1/2} 
\le \left(2C_p^2\sum_{k=M}^{N/2} k^{-2p}\right)^{1/2} \\
&\le \sqrt{2}C_p \left(\int_M^{N/2+1}(x-1)^{-2p} \mathrm{d}x\right)^{1/2}\\
&= \sqrt{\frac{2}{2p-1}}C_p\sqrt{\frac{1}{(M-1)^{2p-1}}-\frac{1}{(N/2)^{2p-1}}} \le \tilde C_p M^{-(2p-1)/2}.
\end{align*}
Here, $C_p$ and $\tilde C_p$ depend on $L,p$ as well as the function $f$, but are independent of $N$ and $M$. 

Recalling that $\sqrt{N}/A_N \xrightarrow{N\to \infty} \sqrt{L}/\|f\|_{L^2(0,L)}$, we find that either the RMSE or the L2NS has similar error estimate:
\begin{equation}
\label{appC:erresti1}
\text{Err}_{\text{RMSE}} \le \tilde C_{p} M^{-(2p-1)/2}, \qquad 
\text{Err}_{\text{L2NS}} \le \hat C_{p} M^{-(2p-1)/2},
\end{equation}
where $\hat C_p = \tilde C_p\sup_{N\in\mathbb{N}} \left(\sqrt{N}/A_N\right)$. To guarantee that the RMSE (or L2NS) is upper bound by a given error bound $\varepsilon$, it is sufficiently to choose $M = \left(\tilde C_p/\varepsilon\right)^{2/(2p-1)}$ (or $M = \left(\hat C_p/\varepsilon\right)^{2/(2p-1)}$). The pre-constant can be estimated in detail using Lemma \ref{appC:lem1} provided that we have the a priori information of $f$. 
\begin{remark}[Agreement between numerical examples and theoretical estimates]
\label{appC:rem2}
The theoretical analysis justifies the orders of the numerical examples in Fig.~\ref{sec3:fig1} of Sect.~\ref{subsec:3-1}. 
For the quadratic function $f_1$, the function is smooth and satisfies $f_1(0)=f_1(L)$. However, the derivatives at endpoints are different: $-L=f^{(1)}(0)\not=f^{(1)}(L)=L$. According to Remark \ref{appC:rem1}, we have $p=2$ and this yields the (sharp) order $O(M^{-3/2})$ as observed. 
For the second case of a linear combination of two Gaussian functions $f_2$, we find that it is smooth. Note that the boundary values are not equal, which indicates $p=1$. This leads to the (sharp) order $O(M^{-1/2})$ for large $M$. On the other hand, we notice that $|f_2^{(q)}(0)-f_2^{(q)}(L)|$ is extremely small for $q=0,1,\ldots$. According to Lemma \ref{appC:lem1}, this implies the error contribution for small $q$ is almost negligible, and hence, the RMSE decays much more rapidly, that is, $O(M^{-11})$ in the right subplot of Fig.~\ref{sec3:fig1}, until a small error is achieved. 
\end{remark}

\subsection{Evaluation of number of repetitions}
\label{subsec:C-2}

When a quantum state: 
$$
\ket{\psi} = \sum_{k=0}^{N-1} \psi_k \ket{k}_n 
$$
is prepared on a quantum computer, we repeat the $Z$-basis measurements of the $n$ qubits for $N_{\text{shot}}$ times. 
For each $k=0,\ldots,N-1$, we can regard the number of successes obtaining the base $\ket{k}_n$ in a sequence of $N_{\text{shot}}$ experiments as a random variable $Z_k$. Note that the probability of success, obtaining the base $\ket{k}_n$, is $p_k = |\psi_k|^2$, and the probability of failure is $1-p_k$. 
The random variable follows binomial distribution with parameters $N_{\text{shot}}\in \mathbb{N}$ and $p_k\in [0,1]$, i.e., $Z_k\sim B(N_{\text{shot}},p_k)$. The mean and the variance are known as $\mathbb{E}[Z_k] = N_{\text{shot}}p_k$ and $\text{Var}(Z_k) = N_{\text{shot}}p_k(1-p_k)$. By the Chebyshev inequality, we have 
\begin{align*}
\mathbb{P}\left(|Z_k-N_{\text{shot}}p_k| \ge \beta\sqrt{N_{\text{shot}}p_k(1-p_k)}\right) \le 1/\beta^2,
\end{align*}
and hence,
\begin{align*}
\mathbb{P}\left(|Z_k/N_{\text{shot}}-p_k| \ge \beta\sqrt{p_k(1-p_k)/N_{\text{shot}}}\right) \le 1/\beta^2,
\end{align*}
for $\beta>0$. This implies that for a sufficiently large fixed constant $\beta\ge 2$, the probability that $|Z_k/N_{\text{shot}}-p_k|\le \beta\sqrt{p_k(1-p_k)/N_{\text{shot}}}$ is close to one. 
Henceforth, we let $\tilde p_k$ be the estimation of $p_k$ and roughly speaking, we have 
$$
|\tilde p_k-p_k| \le \beta\sqrt{p_k(1-p_k)/N_{\text{shot}}},
$$
for sufficiently large $\beta>0$. 

\noindent \underline{\bf Real space readout (RSR) method}\ For the RSR method, $\psi_k = f(x_k)/A_N$, and $p_k=|f(x_k)|^2/A_N^2$ for $k=0,\ldots,N-1$. Then, the approximations is as follows:
$$
f(x_j) \approx A_N \sqrt{\tilde p_j}, \quad f(x_j)/A_N \approx \sqrt{\tilde p_j}, \quad j=0,\ldots,N-1.
$$
Employing the equality $|a-b| = \frac{|a^2-b^2|}{|a+b|}$ for $a,b\ge 0$, we estimate
\begin{align*}
\sum_{j=0}^{N-1} \left|f(x_j)-A_N \sqrt{\tilde p_j}\right|^2 
= \sum_{j=0}^{N-1} \frac{A_N^4\left(|f(x_j)|^2/A_N^2-\tilde p_j\right)^2}{|f(x_j)+A_N\sqrt{\tilde p_j}|^2} 
\le \sum_{j=0}^{N-1} \frac{A_N^4\beta^2p_j(1-p_j)}{N_{\text{shot}}|f(x_j)|^2}
= \frac{\beta^2 A_N^2 (N-1)}{N_\text{shot}}.
\end{align*}
Thus, the RMSE and the L2NS can be estimated:
\begin{align*}
& \text{Err}_{\text{RMSE}} = \left(\frac{1}{N}\sum_{j=0}^{N-1}|f(x_j)-A_N \sqrt{\tilde p_j}|^2\right)^{1/2} \le \frac{\beta A_N}{\sqrt{N_{\text{shot}}}},\\
& \text{Err}_{\text{L2NS}} = \left(\sum_{j=0}^{N-1}\left|\frac{f(x_j)}{A_N}-\sqrt{\tilde p_j}\right|^2\right)^{1/2} \le \frac{\beta \sqrt{N}}{\sqrt{N_{\text{shot}}}}.
\end{align*}
Recalling that the normalized factor $A_N=O(\sqrt{N})$. For both the above errors, we have the order $O(\sqrt{N/N_{\text{shot}}})$. This implies that the order of the number of repetitions is $N_{\text{shot}} = O\left(N(1/\varepsilon)^{2}\right)$ for the RSR method.  

\noindent \underline{\bf Fourier space readout (FSR) method}\ For the FSR method, $\psi_k=c_{k,\text{q}}$, and $p_k = |c_{k,\text{q}}|^2$ for $k=0,\ldots,N-1$. Here, $c_{k,\text{q}}$ is the quantum Fourier coefficient (see Eq.~\eqref{appC:eq-qfc}). In this subsection, we use the notation $c_k$ for simplicity if there is no confusion. According to Eq.~\eqref{appC:eq-arecf}, the approximation is as follows:
$$
f(x_j) \approx \tilde f_{j,M} := \frac{A_N}{\sqrt{N}} \left(\sum_{k=-(M-1)}^{M-1}\mathrm{sgn}(c_k)\sqrt{\tilde p_k}\mathrm{exp}\left(\mathrm{i}2\pi kj/N\right)\right), \quad j=0,\ldots,N-1.
$$
Here, $\tilde p_k := \tilde p_{-k}$ for $k=-(M-1),\ldots,-1$, and we assume that the signs are correctly determined for $k=-(M-1),\ldots,(M-1)$ for simplicity. 
Then, using the triangle inequality, we estimate
\begin{align*}
&\quad \sum_{j=0}^{N-1} \left|f(x_j)-\tilde f_{j,M}\right|^2 \\
&\le \frac{2A_N^2}{N} \sum_{j=0}^{N-1} \left(\left|\sum_{k\in \mathcal{K}}c_k \mathrm{exp}\left(\mathrm{i}2\pi kj/N\right)\right|^2 + \left|\sum_{k=-(M-1)}^{M-1} \mathrm{sgn}(c_k)\left(\sqrt{p_k}-\sqrt{\tilde p_k}\right)\mathrm{exp}\left(\mathrm{i}2\pi kj/N\right)\right|^2\right) \\
&= 2A_N^2\sum_{k\in \mathcal{K}} |c_k|^2 + 2A_N^2\sum_{k=-(M-1)}^{M-1}\left|\sqrt{p_k}-\sqrt{\tilde p_k}\right|^2 \\
&\le 4A_N^2\left(\sum_{k=M}^{N/2} |c_k|^2 + \frac{\beta^2 M}{N_{\text{shot}}}\right).
\end{align*}
Here, $\mathcal{K} := \{-N/2,-N/2+1\ldots,-M, M,M+1,\ldots,N/2-1\}$. By the discussion in Appendix \ref{subsec:C-1}, we have 
$$
\sum_{k=M}^{N/2} |c_k|^2 \le C_p M^{-(2p-1)},
$$
for some integer $p\ge 1$ and some constant $C_p>0$. Thus, the RMSE and the L2NS can be estimated:
\begin{align}
\label{appC:eq-esRMSE}
& \text{Err}_{\text{RMSE}} = \left(\frac{1}{N}\sum_{j=0}^{N-1}|f(x_j)-\tilde f_{j,M}|^2\right)^{1/2} \le \frac{2A_N}{\sqrt{N}}\sqrt{C_p M^{-(2p-1)} + \frac{\beta^2 M}{N_{\text{shot}}}},\\
\label{appC:eq-esL2NS}
& \text{Err}_{\text{L2NS}} = \left(\sum_{j=0}^{N-1}\left|\frac{f(x_j)}{A_N}-\frac{\tilde f_{j,M}}{A_N}\right|^2\right)^{1/2} 
\le 2\sqrt{C_p M^{-(2p-1)} + \frac{\beta^2 M}{N_{\text{shot}}}}.
\end{align}
To achieve the error bound $\varepsilon$, we choose $M=O\left((1/\varepsilon)^{2/(2p-1)}\right)$ and $N_{\text{shot}}=O\left(M(1/\varepsilon)^2\right)=O\left((1/\varepsilon)^{2+2/(2p-1)}\right)$. 
We emphasize that the number of repetitions for the FSR method has no dependence on the grid number $N$, which is a crucial improvement over the RSR method for large grid numbers. 
We mention that the above estimate for the number of repetitions may not be sharp, especially the pre-factor. One should borrow a rigorous discussion from the probability theory for a sharp estimate. Whereas, we conjecture that the orders discussed in this section is optimal, which are confirmed by the numerical examples in Sect.~\ref{subsec:3-2}. 
We end this section with a remark on the FSR method without knowing the parameter $M$. 
\begin{remark}[FSR method without knowledge of approximation parameter]
\label{appC:rem3}
Assume that we obtain the approximation of the quantum Fourier coefficient $c_k$ for each $k=0,\ldots,N-1$, and the sign of $c_k$ is correctly determined. Then, only the sampling error remains. We define 
$$
\tilde f_{j} := \frac{A_N}{\sqrt{N}} \left(\sum_{k=-N/2}^{N/2-1}\mathrm{sgn}(c_k)\sqrt{\tilde p_k}\mathrm{exp}\left(\mathrm{i}2\pi kj/N\right)\right), \quad j=0,\ldots,N-1,
$$
and obtain the estimate for the L2NS as follows (the RMSE is similar):
\begin{align*}
\mathrm{Err}_{\mathrm{L2NS}} &= \left(\sum_{j=0}^{N-1}\left|\frac{f(x_j)}{A_N}-\frac{\tilde f_{j}}{A_N}\right|^2\right)^{1/2} 
= \left(\sum_{k=-N/2}^{N/2-1}\left|\sqrt{p_k}-\sqrt{\tilde p_k}\right|^2\right)^{1/2} = \left(\sum_{k=-N/2}^{N/2-1}\left|\frac{p_k-\tilde p_k}{\sqrt{p_k}+\sqrt{\tilde p_k}}\right|^2\right)^{1/2}\\
&\le \left(\sum_{k=-N/2}^{N/2-1}\left|p_k-\tilde p_k\right|^2/p_k\right)^{1/2} \le \frac{\beta\sqrt{N}}{\sqrt{N_{\mathrm{shot}}}}.
\end{align*}
The last inequality above is a rough estimate, which yields a possible overhead of $O(\sqrt{N})$ that is similar to the case for the RSR method. However, in practice, according to the decay of the Fourier coefficient $|c_k|=O(k^{-p})$, the estimation $\tilde p_k$ will be zero with probability of almost one for any $k\ge O\left(N_{\mathrm{shot}}^{1/2p}\right)$.
This indicates that for a given $N_{\mathrm{shot}}$ (relatively small compared to $N$), the FSR method without a known approximation parameter has a similar behavior as the FSR method with an adaptive approximation parameter $M=O\left(N_{\mathrm{shot}}^{1/2p}\right)$. Putting this $M$ into Eqs.~\eqref{appC:eq-esRMSE} and \eqref{appC:eq-esL2NS} yields the same order of the number of repetitions $N_{\mathrm{shot}}=O\left((1/\varepsilon)^{2+2/(2p-1)}\right)$ as the case for the FSR method with approximation, where $p\ge 1$ depends on the property of the function $f$.  
\end{remark}

\section{Comparison between RSR method and FSR method for a small repetition number}
\label{sec:appD}

\begin{figure}
\centering
\resizebox{14cm}{!}{
\includegraphics[keepaspectratio]{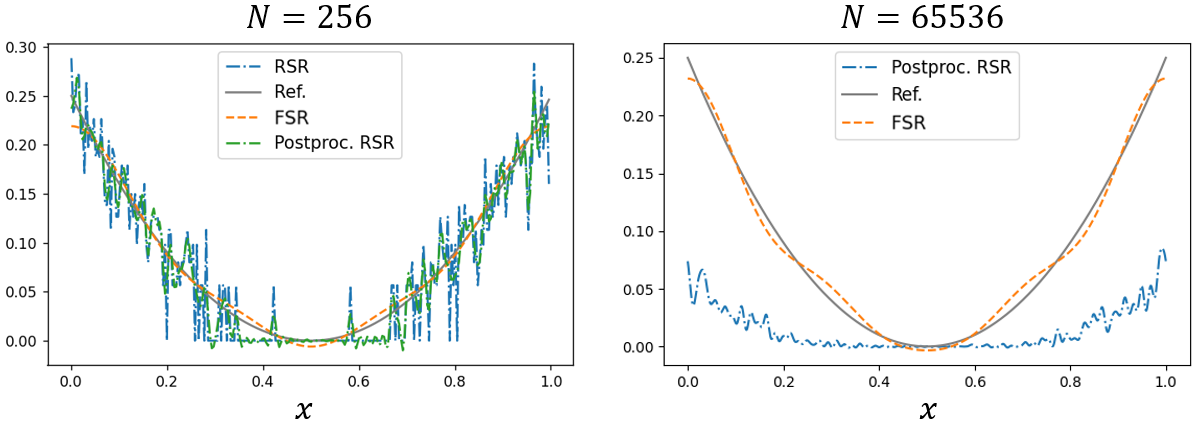}
}
\caption{Reconstructions by the RSR method and the FSR method for a quadratic function. The number of repetitions is chosen relatively small as $N_{\text{shot}}=1000$, and the RSR method with post-processing to reduce the oscillations is also provided. }
\label{appD:fig1}
\end{figure}

\begin{figure}
\centering
\resizebox{14cm}{!}{
\includegraphics[keepaspectratio]{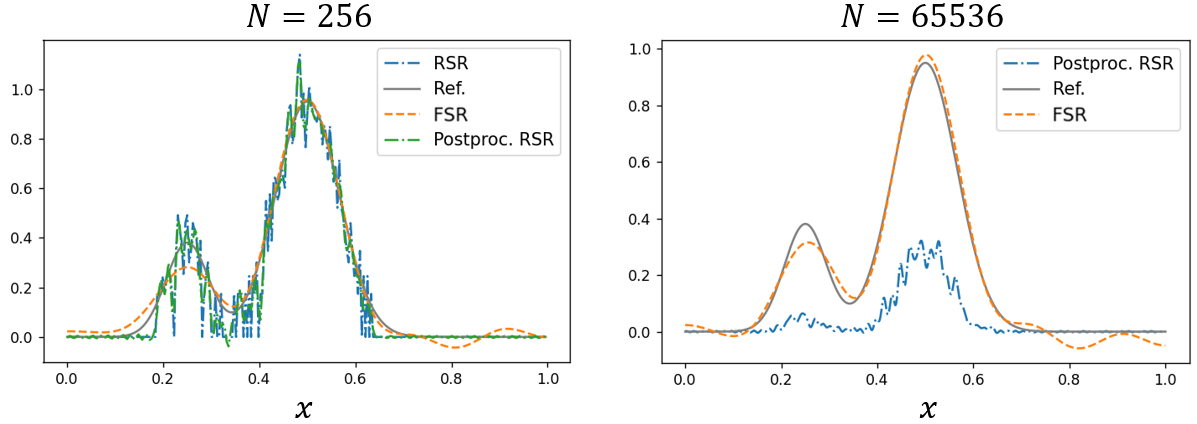}
}
\caption{Reconstructions by the RSR method and the FSR method for a linear combination of Gaussian functions. The number of repetitions is chosen relatively small as $N_{\text{shot}}=1000$, and the RSR method with post-processing to reduce the oscillations is also provided. }
\label{appD:fig2}
\end{figure}

\begin{figure}
\centering
\resizebox{14cm}{!}{
\includegraphics[keepaspectratio]{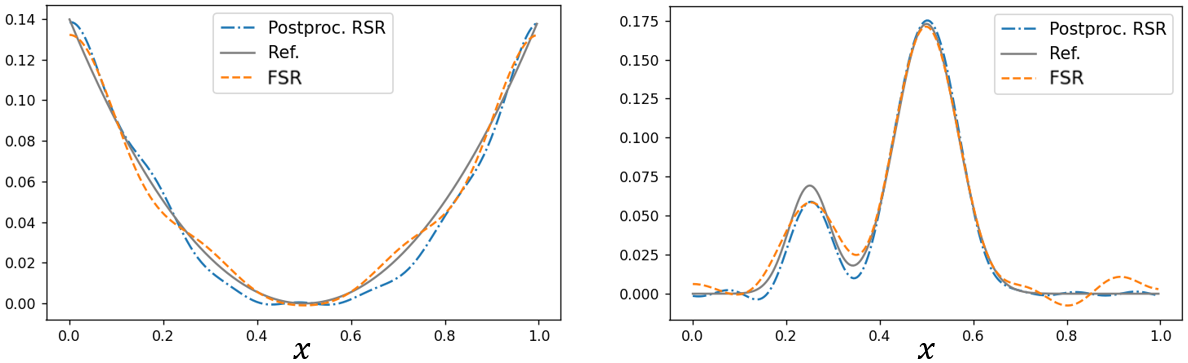}
}
\caption{Reconstructions of the normalized functions/states by the RSR method and the FSR method for both a quadratic function and a linear combination of Gaussian functions. Here, the grid number is $N=256$, and the number of repetitions is chosen relatively small as $N_{\text{shot}}=1000$. Moreover, the results of the post-processed RSR are further smoothened using a smaller truncation number $8$.}
\label{appD:fig3}
\end{figure}

\begin{figure}[ht]
\centering
\resizebox{14cm}{!}{
\includegraphics[keepaspectratio]{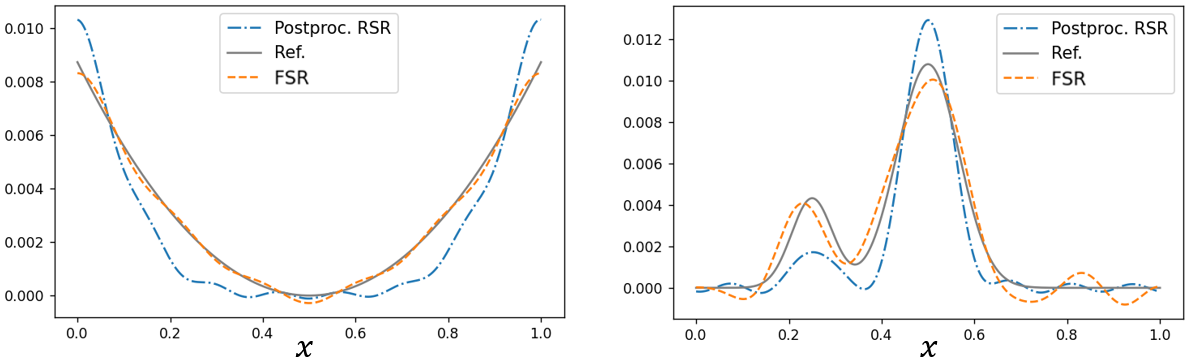}
}
\caption{Reconstructions of the normalized functions/states by the RSR method and the FSR method for both a quadratic function and a linear combination of Gaussian functions. Here, the grid number is $N=65536$, and the number of repetitions is chosen relatively small as $N_{\text{shot}}=1000$. Moreover, the results of the post-processed RSR are further smoothened using a smaller truncation number $8$.}
\label{appD:fig4}
\end{figure}

For the practical interest of applications, we give the plots with limited number of repetitions $N_\text{shot}=1000$. The reconstructions, with a small grid number $N=256$ and a large grid number $N=65536$, of the quadratic function and the linear combination of Gaussian functions discussed in Sect.~\ref{sec3} are illustrated in Figs.~\ref{appD:fig1} and \ref{appD:fig2}, respectively. 
The results show that the FSR method could provide roughly the shape of the function even for a small number of repetition $N_{\text{shot}}=1000$. For a small grid number $N=256$, the RSR method also gives the shape although the oscillation should be further reduced by suitable post-processing method. On the other hand, for a large grid number $N=65536$, the norm information after the post-processing is lost. 
Besides, if we consider the normalized reconstructions, then the results for the post-processed RSR method look better than the ones without normalization, see Figs.~\ref{appD:fig3} and \ref{appD:fig4}. Here, we further remove the high frequency components after $8$ for the post-processed RSR method to derive relatively smooth reconstructions. 
The numerical plots imply that for small grid numbers, the FSR method does not change a lot from the RSR method with a suitable post-processing. However, for large grid numbers, even the oscillations can be removed by a good post-processing, the error is much larger than the FSR method under the same number of repetitions.  

\section{Fully quantum Fourier space readout (fQFSR) method}
\label{sec:appE}

Different from the FSR method, in this appendix, we provide the quantum circuit for the overlap-based fully quantum Fourier space readout method by the swap test. 
The aim is to obtain the overlap mentioned in Sect.~\ref{subsec:4-1}. Instead of the swap test, one can alternatively use the Hadamard (or switch) test or the interferometric measurement protocol \cite{Wu.2025pre,Kyriienko2020,Scali.2024}. Here, we adopt the swap test for simplicity, and we intend to discuss the crucial difference between the quantum circuit without the approximation and the one with an approximation. 

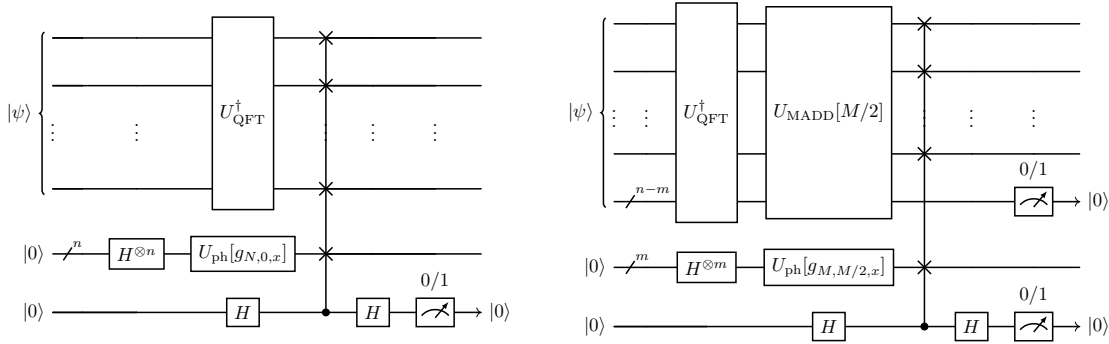
\begin{figure}
\centering
\resizebox{15cm}{!}{
\begin{quantikz}
\lstick[4]{\ket{\psi}} & \qw &  & \gate[4]{U_{\text{QFT}}^\dag} & \targX{} & \qw & \qw & \\[-0cm]
 & \qw &  &  & \targX{} & \qw & \qw & \\[-0cm]
\setwiretype{n}\vdots &  & \vdots & \vdots & \vdots & \vdots & \vdots & \\
 & \qw &  &  & \targX{} & \qw & \qw & \\
\lstick{\ket{0}} & \qwbundle{n} & \gate{H^{\otimes n}} & \gate{U_{\text{ph}}[g_{N,0,x}]} & \targX{} & \qw & \qw & \\
\lstick{\ket{0}} & \qw & \qw & \gate{H} & \ctrl{-5} & \gate{H} & \meter{0/1}\arrow[r] & \rstick{\ket{0}} 
\end{quantikz}
\qquad
\begin{quantikz}
\lstick[5]{\ket{\psi}} &  & \gate[5]{U_{\text{QFT}}^\dag} & \gate[5]{U_{\text{MADD}}[M/2]} & \targX{} &  &  &  \\[-0cm]
 &  &  &  & \targX{} &  &  &  \\[-0cm]
\setwiretype{n}\vdots & \vdots & \vdots &  & \vdots & \vdots & \vdots & \\
 &  &  &  & \targX{} &  &  &  \\[-0cm]
 & \qwbundle{n-m} &  &  &  &  & 
\meter{0/1}\arrow[r] & \rstick{\ket{0}} \\
\lstick{\ket{0}} & \qwbundle{m} & \gate{H^{\otimes m}} & \gate{U_{\text{ph}}[g_{M,M/2,x}]} & \targX{} &  &  &  \\
\lstick{\ket{0}} & \qw & \qw & \gate{H} & \ctrl{-6} & \gate{H} & \meter{0/1}\arrow[r] & \rstick{\ket{0}}
\end{quantikz}
}
\caption{Quantum circuits for calculating the value of the $j$-th real-grid point by the swap test. The left subplot is for the precise fQFSR method, while the right subplot is for the approximate one with an approximation parameter $M=2^m$ for some integer $m< n$. The polynomial phase gate $U_{\text{ph}}[g_{K,K_0,x}]$ is defined by Eq.~\eqref{appE:eq-ph}, and the shift of the basis is implemented by a modular adder defined by Eq.~\eqref{appE:eq-add}. }
\label{appE:fig1}
\end{figure}

First, we discuss the quantum circuit without the approximation which is given in the left subplot in Fig.~\ref{appE:fig1}. Here, the polynomial phase gate $U_{\text{ph}}[g_{K,K_0,x}]$, $x\in [0,L]$ is defined by
\begin{equation}
\label{appE:eq-ph}
U_{\text{ph}}[g_{K,K_0,x}] := \sum_{k=0}^{K-1} \mathrm{exp}\left(-\mathrm{i}2\pi (k-K_0)x/L\right)\ket{k}\bra{k}.
\end{equation}
Using the swap test, we find that the probability of post-selecting the last qubit to be $\ket{0}$ is
$$
(1+|\langle \phi_x|U_{\text{QFT}}^\dag \ket{\psi}|^2)/2,
$$
with which we can estimate the overlap $|\langle \phi_x|U_{\text{QFT}}^\dag \ket{\psi}|$. Here, $\ket{\phi_x}$ is the quantum state defined by
$$
\ket{\phi_x} := \frac{1}{\sqrt{N}}\sum_{k=0}^{N-1} \mathrm{exp}\left(-\mathrm{i}2\pi kx/L\right)\ket{k}, \quad x\in [0,L].
$$
By the discussions in Appendix \ref{subsec:C-1}, we find that for $x=x_j:= jL/N$, we have the equality:
$$
\langle \phi_{x_j} |U_{\text{QFT}}^{\dag}\ket{\psi} = \frac{1}{\sqrt{N}} \sum_{k=0}^{N-1} c_{k,\text{q}} \mathrm{exp}\left(\mathrm{i}2\pi kj/N\right) = \frac{f(x_j)}{A_N}, \quad j=0,\ldots,N-1.
$$
Moreover, for $x\not\in \{jL/N, j=0,\ldots,N-1\}$, the last equality above becomes an approximation. In other words, 
$$
A_N \langle \phi_x |U_{\text{QFT}}^{\dag}\ket{\psi} \xrightarrow{N\to \infty} f(x).
$$
Noting that $A_N \propto \sqrt{N}$, to reconstruct $f$ with the error bound $\varepsilon$, we need to estimate the overlap $\langle \phi_x |U_{\text{QFT}}^{\dag}\ket{\psi}$ with an error bound $\varepsilon/\sqrt{N}$. According to the estimation of the stochastic error, the number of repetitions should be proportional to $N/\varepsilon^{2}$. This indicates that such a readout method is not efficient for large grid numbers. 

Next, we discuss the quantum circuit with a given approximation parameter $M=2^m < N$. Note that the quantum Fourier coefficients satisfy $c_{k,\text{q}} = \overline{c_{N-k,\text{q}}}$ for $k=1,\ldots,N/2$ provided that $f$ is real-valued. We need a quantum modular adder $U_{\text{MADD}}[M/2]$ to make the $M$ most dominant Fourier coefficients appear in the amplitudes of the first $M$ bases. The quantum modular adder is defined by
\begin{equation}
\label{appE:eq-add}
U_{\text{MADD}}[M/2] \ket{k}_n := \ket{k+M/2}_n, \quad k=0,\ldots,N-1,
\end{equation}
which can be implemented using the QFTs without any ancillary qubits (gate complexity $O(n^2)$) \cite{Yuan.2023} or the Toffoli gates and CNOT gates with at most $n$ ancillary qubits (gate complexity $O(n)$) \cite{Li.2020}. The quantum state after applying the quantum modular adder is 
\begin{align*}
\ket{\hat \psi}_n &= \sum_{k=0}^{N-1} c_{k,\text{q}} \ket{k+M/2}_n 
= \sum_{k=M/2}^{N-1} c_{k-M/2,\text{q}} \ket{k}_n + \sum_{k=0}^{M/2-1} c_{k+N-M/2,\text{q}} \ket{k}_n \\
&= \sum_{k=M/2}^{N-1} c_{k-M/2,\text{q}} \ket{k}_n + \sum_{k=0}^{M/2-1} \overline{c_{M/2-k,\text{q}}} \ket{k}_n = \sum_{k=0}^{N-1} c_{k-M/2,\text{q}} \ket{k}_n.
\end{align*}
Here, we naturally extend the quantum Fourier coefficients for $k=-M/2,-M/2+1,\ldots,-1$ by the definition \eqref{appC:eq-qfc}, and we have $c_{k,\text{q}}=\overline{c_{-k,\text{q}}}$. 
Using the swap test in the right subplot in Fig.~\ref{appE:fig1}, we find that the probability of post-selecting the last $n-m$ qubits of the input register and the last qubit to be $\ket{0}_{n-m+1}$ is
$$
\frac{C_M}{2}\left(1 + \left|\Big\langle\phi_x^M\ket{\hat\psi^M}\right|^2\right),
$$
where
\begin{align*}
&C_M := \sum_{k=0}^{M-1} |c_{k-M/2,\text{q}}|^2, \\
&\ket{\hat\psi^M} := \frac{1}{\sqrt{C_M}} \sum_{k=0}^{M-1} c_{k-M/2,\text{q}} \ket{k}_m, \\
&\ket{\phi_x^M} := \frac{1}{\sqrt{M}} \sum_{k=0}^{M-1} \mathrm{exp}\left(-\mathrm{i}2\pi (k-M/2)x/L\right)\ket{k}_m, \quad x\in [0,L].
\end{align*}
Here, $C_M = 1 - \sum_{k=M/2}^{N-M/2-1} |c_{k,\text{q}}|^2$ is close to $1$ if $M$ is large enough, and it can be approximately estimated by the probability of post-selecting the last $n-m$ qubits of the input register to be $\ket{0}_{n-m}$. Therefore, we have the approximation:
$$
f(x) \approx \frac{A_N}{\sqrt{N}} \sum_{k=-M/2}^{M/2-1} c_{k,\text{q}} \mathrm{exp}\left(\mathrm{i}2\pi kx/L\right)
= \frac{A_N \sqrt{M C_M}}{\sqrt{N}}\left|\Big\langle\phi_x^M\ket{\hat\psi^M}\right|.
$$
Different from the quantum circuit without the approximation, the number of repetitions is now proportional to $A_N^2 M C_M/(N\varepsilon^2) \propto M/\varepsilon^2$. According to the discussion of the decay rate for the Fourier coefficients in Appendix \ref{subsec:C-1}, we have $M=O(1/\varepsilon^{2/(2p-1)})$ for some positive integer $p$. This gives the same order as the FSR method proposed in the main manuscript. 

Finally, we emphasize the difference between the FSR method and the above overlap-based fQFSR method with an approximation. One key difference is that the approximation parameter of the FSR method can be automatically determined corresponding to the number of repetitions, while the approximation parameter for the fQFSR method should be suitably given in advance. Another importance difference is the balance of required quantum resources and classical resources. The fQFSR method uses limited classical computations, so that the required quantum resources are proportional to the number of target points. 
On the other hand, the hybrid method FSR enables us to minimize the quantum computations to derive only the features (i.e., dominant Fourier coefficients) via quantum computers and recover the values at the target points on classical computers (possibly using parallel computing). For a relatively large number of target points, this is much more cheaper in the early stage of the quantum computing. 

\end{document}